\PassOptionsToPackage{unicode}{hyperref}
\PassOptionsToPackage{hyphens}{url}
\PassOptionsToPackage{dvipsnames,svgnames,x11names}{xcolor}
\documentclass[
]{article}

\usepackage{amsmath,amssymb}
\usepackage{iftex}
\ifPDFTeX
  \usepackage[T1]{fontenc}
  \usepackage[utf8]{inputenc}
  \usepackage{textcomp} 
\else 
  \usepackage{unicode-math}
  \defaultfontfeatures{Scale=MatchLowercase}
  \defaultfontfeatures[\rmfamily]{Ligatures=TeX,Scale=1}
\fi
\usepackage{lmodern}
\ifPDFTeX\else  
\fi
\IfFileExists{upquote.sty}{\usepackage{upquote}}{}
\IfFileExists{microtype.sty}{
  \usepackage[]{microtype}
  \UseMicrotypeSet[protrusion]{basicmath} 
}{}
\makeatletter
\@ifundefined{KOMAClassName}{
  \IfFileExists{parskip.sty}{%
    \usepackage{parskip}
  }{
    \setlength{\parindent}{0pt}
    \setlength{\parskip}{6pt plus 2pt minus 1pt}}
}{
  \KOMAoptions{parskip=half}}
\makeatother
\usepackage{xcolor}
\ifx\paragraph\undefined\else
  \let\oldparagraph\paragraph
  \renewcommand{\paragraph}[1]{\oldparagraph{#1}\mbox{}}
\fi
\ifx\subparagraph\undefined\else
  \let\oldsubparagraph\subparagraph
  \renewcommand{\subparagraph}[1]{\oldsubparagraph{#1}\mbox{}}
\fi

\usepackage{longtable,booktabs,array}
\usepackage{calc} 
\usepackage{etoolbox}
\makeatletter
\patchcmd\longtable{\par}{\if@noskipsec\mbox{}\fi\par}{}{}
\makeatother
\IfFileExists{footnotehyper.sty}{\usepackage{footnotehyper}}{\usepackage{footnote}}
\makesavenoteenv{longtable}
\usepackage{graphicx}
\makeatletter
\def\maxwidth{\ifdim\Gin@nat@width>\linewidth\linewidth\else\Gin@nat@width\fi}
\def\maxheight{\ifdim\Gin@nat@height>\textheight\textheight\else\Gin@nat@height\fi}
\makeatother
\setkeys{Gin}{width=\maxwidth,height=\maxheight,keepaspectratio}
\makeatletter
\def\fps@figure{htbp}
\makeatother
\newlength{\cslhangindent}
\newlength{\csllabelwidth}
\newlength{\cslentryspacingunit} 
\newenvironment{CSLReferences}[2] 
 {
  \setlength{\parindent}{0pt}
  \ifodd #1
  \let\oldpar\par
  \def\par{\hangindent=\cslhangindent\oldpar}
  \fi
  \setlength{\parskip}{#2\cslentryspacingunit}
 }%
 {}
\usepackage{calc}

\usepackage{booktabs}
\usepackage{longtable}
\usepackage{array}
\usepackage{multirow}
\usepackage{wrapfig}
\usepackage{float}
\usepackage{colortbl}
\usepackage{pdflscape}
\usepackage{tabu}
\usepackage{threeparttable}
\usepackage{threeparttablex}
\usepackage[normalem]{ulem}
\usepackage{makecell}
\usepackage{xcolor}
\usepackage{arxiv}
\usepackage{orcidlink}
\usepackage{amsmath}
\usepackage[T1]{fontenc}
\makeatletter
\@ifpackageloaded{caption}{}{\usepackage{caption}}
\AtBeginDocument{%
\ifdefined\contentsname
  \renewcommand*\contentsname{Table of contents}
\else
  \newcommand\contentsname{Table of contents}
\fi
\ifdefined\listfigurename
  \renewcommand*\listfigurename{List of Figures}
\else
  \newcommand\listfigurename{List of Figures}
\fi
\ifdefined\listtablename
  \renewcommand*\listtablename{List of Tables}
\else
  \newcommand\listtablename{List of Tables}
\fi
\ifdefined\figurename
  \renewcommand*\figurename{Figure}
\else
  \newcommand\figurename{Figure}
\fi
\ifdefined\tablename
  \renewcommand*\tablename{Table}
\else
  \newcommand\tablename{Table}
\fi
}
\@ifpackageloaded{float}{}{\usepackage{float}}
\floatstyle{ruled}
\@ifundefined{c@chapter}{\newfloat{codelisting}{h}{lop}}{\newfloat{codelisting}{h}{lop}[chapter]}
\floatname{codelisting}{Listing}

\makeatother
\makeatletter
\@ifpackageloaded{caption}{}{\usepackage{caption}}
\@ifpackageloaded{subcaption}{}{\usepackage{subcaption}}
\makeatother
\makeatletter
\@ifpackageloaded{tcolorbox}{}{\usepackage[skins,breakable]{tcolorbox}}
\makeatother
\makeatletter
\@ifundefined{shadecolor}{\definecolor{shadecolor}{rgb}{.97, .97, .97}}
\makeatother
\ifLuaTeX
  \usepackage{selnolig}  
\fi
\IfFileExists{bookmark.sty}{\usepackage{bookmark}}{\usepackage{hyperref}}
\IfFileExists{xurl.sty}{\usepackage{xurl}}{} 
\hypersetup{
  pdftitle={Mapping historical forest biomass for stock-change assessments at parcel to landscape scales},
  pdfauthor={Lucas K Johnson; Michael J Mahoney; Madeleine L Desrochers; Colin M Beier},
  pdfkeywords={Landsat, LiDAR, aboveground biomass, machine
learning, land cover},
  colorlinks=true,
  linkcolor={blue},
  filecolor={Maroon},
  citecolor={Blue},
  urlcolor={Blue},
  pdfcreator={LaTeX via pandoc}}

\renewcommand{\today}{2023-04-04}
\newcommand{\runninghead}{A Preprint }
\renewcommand{\runninghead}{Mapping forest biomass for stock-change }
\title{Mapping historical forest biomass for stock-change assessments at
parcel to landscape scales}
\author{
\textbf{Lucas K Johnson}~\orcidlink{0000-0002-7953-0260}\\Graduate
Program in Environmental Science\\State University of New York College
of Environmental Science and
Forestry\\Syracuse,\ 13210\\\href{mailto:ljohns11@esf.edu}{ljohns11@esf.edu}\\\\\\
\textbf{Michael J Mahoney}~\orcidlink{0000-0003-2402-304X}\\Graduate
Program in Environmental Science\\State University of New York College
of Environmental Science and
Forestry\\Syracuse,\ 13210\\\href{mailto:mjmahone@esf.edu}{mjmahone@esf.edu}\\\\\\
\textbf{Madeleine L
Desrochers}~\orcidlink{0000-0003-4194-3577}\\Department of Sustainable
Resources Management\\State University of New York College of
Environmental Science and
Forestry\\Syracuse,\ 13210\\\href{mailto:mldesroc@syr.edu}{mldesroc@syr.edu}\\\\\\
\textbf{Colin M Beier}~\orcidlink{0000-0003-2692-7296}\\Department of
Sustainable Resources Management\\State University of New York College
of Environmental Science and
Forestry\\Syracuse,\ 13210\\\href{mailto:cbeier@esf.edu}{cbeier@esf.edu}}
\date{2023-04-04}
\begin{document}
\maketitle
\begin{abstract}
Understanding historical forest dynamics, specifically changes in forest
biomass and carbon stocks, has become critical for assessing current
forest climate benefits and projecting future benefits under various
policy, regulatory, and stewardship scenarios. Carbon accounting
frameworks based exclusively on national forest inventories are limited
to broad-scale estimates, but model-based approaches that combine these
inventories with remotely sensed data can yield contiguous
fine-resolution maps of forest biomass and carbon stocks across
landscapes over time. Here we describe a fundamental step in building a
map-based stock-change framework: mapping historical forest biomass at
fine temporal and spatial resolution (annual, 30m) across all of New
York State (USA) from 1990 to 2019, using freely available data and
open-source tools.

Using Landsat imagery, US Forest Service Forest Inventory and Analysis
(FIA) data, and off-the-shelf LiDAR collections we developed three
modeling approaches for mapping historical forest aboveground biomass
(AGB): training on FIA plot-level AGB estimates (direct), training on
LiDAR-derived AGB maps (indirect), and an ensemble averaging predictions
from the direct and indirect models. Model prediction surfaces (maps)
were tested against FIA estimates at multiple scales. All three
approaches produced viable outputs, yet tradeoffs were evident in terms
of model complexity, map accuracy, saturation, and fine-scale pattern
representation. The resulting map products can help identify where,
when, and how forest carbon stocks are changing as a result of both
anthropogenic and natural drivers alike. These products can thus serve
as inputs to a wide range of applications including stock-change
assessments, monitoring reporting and verification frameworks, and
prioritizing parcels for protection or enrollment in improved management
programs.
\end{abstract}
{\bfseries \emph Keywords}
\def\sep{\textbullet\ }
Landsat \sep LiDAR \sep aboveground biomass \sep machine learning \sep 
land cover

\ifdefined\Shaded\renewenvironment{Shaded}{\begin{tcolorbox}[breakable, enhanced, borderline west={3pt}{0pt}{shadecolor}, interior hidden, boxrule=0pt, sharp corners, frame hidden]}{\end{tcolorbox}}\fi

\hypertarget{introduction}{%
\section{Introduction}\label{introduction}}

Forests are among the most effective natural carbon sinks and thus are
essential in stabilizing Earth's climate, but their capacity to provide
this critical service has been strongly shaped by past and present
anthropogenic impacts. Understanding the spatiotemporal dynamics of
forest carbon in relation to human activities has become increasingly
important as policymakers and stakeholders look to nature-based
solutions to reduce atmospheric greenhouse gas (GHG) concentrations and
mitigate climate change (Malmsheimer et al. 2008; Fargione et al. 2018).
With a better grasp of local social and ecological conditions across the
forest landscape, decision-makers could identify and prioritize parcels
of land suitable for different strategies such as reforestation, avoided
conversion, or enhanced forest management, in order to sustain and/or
increase carbon sequestration and effectively offset GHG emissions from
other sectors (Houghton 2005; Houghton et al. 2012). To quantify
potential climate benefits, carbon status and trends are typically
assessed using a stock-change methodology that requires historical data
and ongoing monitoring efforts via permanent plot networks.

National forest inventories (NFI) like the USDA's Forest Inventory and
Analysis (FIA) program provide estimates of forest biomass, carbon
stocks, and stock-changes at large scales based on their extensive
sampling design. Although these programs have offered fundamental
insights and essential data on forest carbon dynamics over the past
three decades (Buendia et al. 2019; Woodall et al. 2015), they are
limited spatially by the sample density and remeasurement frequency
(McRoberts 2011), and thus cannot represent fine-scaled patterns and
dynamics most relevant to planning and decision-making. Model-based
approaches, which combine field data like the FIA with wall-to-wall
remotely-sensed data can fill this need by producing predictions for all
map units (pixels) in a given area.

Largely due to limitations of the available data, implementing
model-based approaches for characterizing historical spatiotemporal
dynamics of forest carbon remains challenging. Remotely-sensed data best
describes the most prominent aboveground components of a forest, and for
this reason aboveground biomass (AGB) often serves as an initial target
variable (Houghton, Hall, and Goetz 2009) before empirical conversions
to specific carbon pools are made (Heath et al. 2009; Woodall et al.
2011). Airborne LiDAR has been established as a highly valuable
remotely-sensed data source for such AGB mapping efforts, but is often
collected for irregularly defined boundaries at local to regional
scales, resulting in spatiotemporal patchworks when pooled together for
broad-scale applications (Johnson et al. 2022; Huang et al. 2019;
Skowronski and Lister 2012). Remotely-sensed optical imagery offers far
better spatial coverage and temporal consistency than airborne LiDAR
point clouds, but cannot characterize forest structure with the same
level of detail nor at the same spatial resolution. Optical datasets
still provide the best set of historical earth surface observations
available; in particular, the Landsat program offering spectral
information at a 30 m resolution for the past four decades has supported
a broad array of historical time series mapping efforts (Hansen and
Loveland 2012; Banskota et al. 2014; Wulder et al. 2022). More recent
spaceborne remote sensing missions that collect LiDAR and synthetic
aperture radar (SAR) may offer benefits for quantifying forest structure
at similarly broad scales, but these platforms cannot match the
historical continuity offered by Landsat (Dubayah et al. 2014; Abdalati
et al. 2010; Torres et al. 2012; Rosenqvist et al. 2007).

A handful of studies have used Landsat time series imagery for
multi-annual, fine-resolution, broad-scale AGB mapping (Kennedy, Ohmann,
et al. 2018; Matasci et al. 2018; Hudak et al. 2020). These efforts can
be categorized into `direct' approaches, where models were fit using AGB
measurements from FIA field plots (Kennedy, Ohmann, et al. 2018), and
`indirect' approaches, where models were fit to AGB predictions from
separate models trained with LiDAR data (Matasci et al. 2018; Hudak et
al. 2020). Direct approaches offer a degree of parsimony relative to
their indirect counterparts, and limit the propagation of errors through
multiple stages of modeling. Indirect approaches could yield more
accurate predictions due to the availability of a larger model training
sample comprised of LiDAR-based predictions (pixels). In theory a sample
of LiDAR-based predictions would cover a wider range of AGB conditions,
have improved geolocation accuracy, and offer better spatial
compatibility with Landsat pixels relative to traditional field plots
(Hudak et al. 2020). These two overarching approaches (direct and
indirect) have only been compared for snapshots in time (single year
mapping), over a relatively small (820,000 ha) and homogenous section of
boreal forest in Alaska (Strunk et al. 2014), as well as over Mexico
with the Mexican NFI and the addition of SAR data (Urbazaev et al.
2018).

In this paper, as part of a broader effort for map-based forest carbon
accounting across New York State (NYS), we present methods for
translating FIA's discrete plot-based inventory to 30 years (1990-2019)
of annual statewide AGB maps at a 30 m resolution. The resulting map
products provide the necessary data to replicate FIA's stock-change
accounting approach in a spatially explicit manner with the flexibility
to produce outputs at scales ranging from individual parcels to the
entire state. The models we developed to achieve these ends demonstrate
what is to our knowledge the first attempt to synthesize direct and
indirect approaches. We used Landsat time series imagery, FIA plots, and
publicly available off-the-shelf LiDAR data to develop an ensemble of
these two distinct modeling strategies (direct and indirect) that
leveraged their relative strengths and improved the predictive accuracy
of our overall approach. We assessed agreement between mapped
predictions from all three approaches (direct, indirect, and ensemble)
and an independent set of FIA estimates across a range of scales. These
methods using publicly available data and open-source tools are
flexible, efficient, and extensible in space and time, thus providing a
framework for those seeking to develop maps of forest AGB dynamics for
both retrospective and monitoring objectives alike. Results produced
following this framework not only provide inputs for stock-change
analyses at scales germane to management, but will also broadly support
forest stewardship, future research, and ongoing planning.

\hypertarget{data-and-methods}{%
\section{Data and methods}\label{data-and-methods}}

\hypertarget{overview}{%
\subsection{Overview}\label{overview}}

\begin{figure}

{\centering \includegraphics{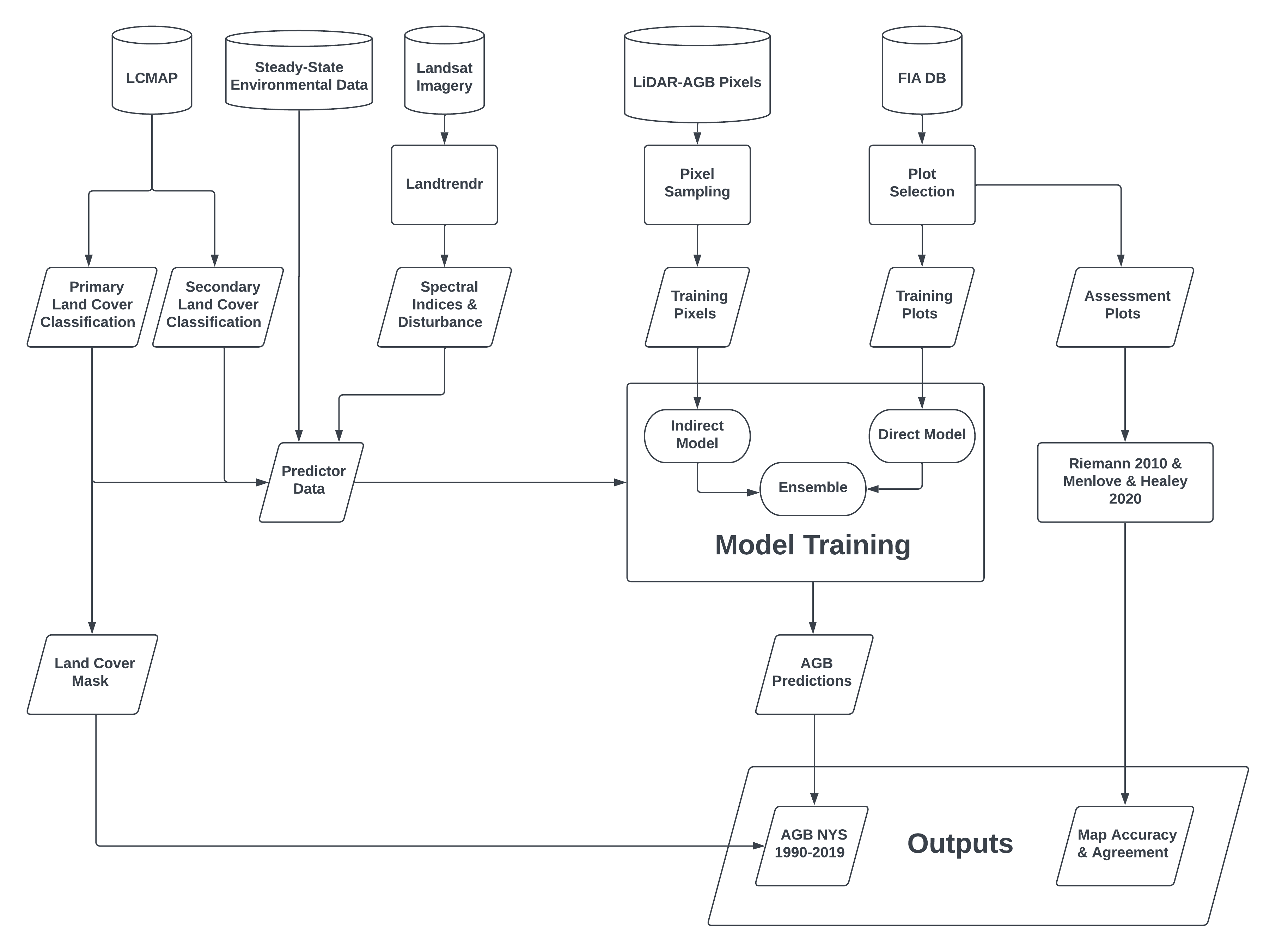}

}

\caption{\label{fig-flowchart}A flowchart diagram showing the key
elements of the modeling and mapping methodology. Cylinders represent
data repositories, parallelograms represent data products and results,
rectangles represent processing steps, and ovals represent models.}

\end{figure}

We developed three modeling approaches (Figure~\ref{fig-flowchart}) to
map aboveground biomass (AGB) annually across New York State (NYS). The
direct approach used AGB estimates at USDA Forest Inventory and Analysis
(FIA; Gray et al. (2012)) field plots as a dependent variable. The
indirect approach used LiDAR-based predictions of AGB developed by
Johnson et al. (2022) as a dependent variable. For both approaches, the
respective dependent variables were associated with predictors derived
from temporally matching Landsat imagery and landcover classifications,
as well as temporally static climate, topographic, and ecological
layers. We used each of these combined datasets to produce separate
stacked ensemble models composed of several machine learning (ML)
models. Predictions from these two approaches were averaged to create a
third ensemble approach. Each of the three modeling approaches were used
to make annual (1990-2019) AGB predictions at a 30m resolution across
the entire state, and the resulting maps were assessed with a common set
of independent FIA plots.

\hypertarget{sec-study-area}{%
\subsection{Study area}\label{sec-study-area}}

NYS covers 141,297 km\textsuperscript{2} in the Northeastern US and was
approximately 59\% forested as of 2019 (USFS 2020). The forests are
dominated by Northern hardwoods-hemlock types but include Appalachian
oak and beech-maple-basswood forests in the western and southern regions
of the state respectively (Dyer 2006). Like much of the US Northeast,
NYS was extensively deforested during the 18th and 19th centuries, with
subsequent reforestation, and conservation resulting in a landscape
dominated by forest stands that are now over 100 years old (Whitney
1994; Lorimer 2001; Michael J. Mahoney et al. 2022). NYS created the
Forest Preserve in 1885, establishing the foundation for what became the
Adirondack and Catskill Parks decades later. Any state-owned or acquired
lands within these parks has since been designated as `forever wild' and
has largely been protected from timber harvesting. More recent land use
dynamics indicate that total agricultural area has continued to decline
in the state and has been replaced by similar extents of forested and
developed lands (Widmann et al. 2012; Widmann 2016). Total forest area
was estimated to have peaked as of 2012 and forest loss due to continued
human development has recently outpaced gains due to agricultural
abandonment (Widmann 2016; USFS 2020). Harvesting activities,
weather-related events, and insect outbreaks drive disturbance and
damage patterns within consistently forested areas (Kosiba et al. 2018;
USFS 2020).

\hypertarget{sec-fia}{%
\subsection{Field data}\label{sec-fia}}

Two field datasets were compiled from the FIA inventory in NYS for the
distinct purposes of model development and map assessment. The FIA
program compiled AGB estimates for trees \(\geq\) 12.7 cm (5 in)
diameter at breast height (Gray et al. 2012), and were converted to
units of megagrams per hectare (Mg ha\textsuperscript{-1}). The FIA uses
permanent inventory plots arranged in a quasi-systematic hexagonal grid
that are divided into five panels, each assumed to have complete spatial
coverage over the state, and remeasured on a 5--7 year basis (Bechtold
and Patterson 2005). Tree measurements, and subsequently AGB estimates
based on allometrics, were only recorded on portions of plots considered
forested. For an area to be considered forested by the FIA, the area
must be at least 10\% stocked with trees, at least 0.4 ha (1 acre) in
size, and at least 36.58 m (120 ft) wide. Any lands meeting these
minimum requirements, but developed for nonforest land uses, were not
considered forested. By this definition, it is likely that some
nonforest conditions contained AGB that was not measured. In absence of
additional information, however, we assumed that any nonforest
conditions represented 0 AGB.

FIA plots are composed of four identical circular subplots with radii of
7.32 m (24 ft), with one subplot centered at the macroplot centroid and
three subplots located 36.6 m (120 ft) away at azimuths of 360°, 120°,
and 240° (Bechtold and Patterson 2005). The plot locations were provided
by the FIA program in the form of average coordinates, collected over
multiple repeat visits, representing the centroid of the center subplot,
which we then used to build a polygon dataset representing the entire
plot layout including all four subplots. Averaged coordinates were
necessary due to the lack of precision of initial GPS coordinates for
the macroplot centroids (Hoppus and Lister 2005; Cooke 2000). We use the
phrase `FIA plot' to refer to the aggregation of all four subplots.

We only considered FIA plots following the national plot design where
all subplots were marked as measured. Importantly, excluding
non-measured plots does not invalidate FIA's probability sample because
the FIA program assumes these plots to be randomly distributed across
the landscape (Bechtold and Patterson 2005). Further, when available
plots were inventoried more than once, single instances were selected
randomly to avoid replication. These initial selection criteria resulted
in a pool of 5,144 plots inventoried between 2002 and 2019. We then
divided this set of plots into the model development and map assessment
datasets using FIA's panel designation, with one of the five panels
randomly selected and all plots with this designation assigned to the
map assessment dataset, and the remaining plots assigned to the model
development dataset. In this way we partitioned 20\% of the available
plot data for an independent map assessment, yielding a probability
sample with complete spatial coverage which we used to generate unbiased
estimates of map agreement metrics (Stehman and Foody 2019; Riemann et
al. 2010).

For the model development dataset we further selected the 1,954
completely forested plots to ensure that non-response in nonforest
conditions would not degrade the relationship between predictors and
plot-level AGB. However, to train and test our models with information
covering the broadest possible range of conditions we added a set of 95
completely nonforested plots that were identified as true zeroes (AGB)
based on LIDAR-derived maximum heights \(\leq\) 1 m (Johnson et al.
2022). The model development dataset contained 2,049 unique plots
(Table~\ref{tbl-plotcount}). For the map assessment dataset we filtered
plots external to our mapped area based on our landcover mask
(Section~\ref{sec-maps}), as these plots were considered outside our
population of interest, resulting in 545 total plots
(Table~\ref{tbl-plotcount}).

\hypertarget{tbl-plotcount}{}
\begin{table}
\caption{\label{tbl-plotcount}Annual counts of FIA plots divided into model development and map
assessment datasets. }\tabularnewline

\centering\begingroup\fontsize{10}{12}\selectfont

\begin{tabular}[t]{lrr}
\toprule
\multicolumn{1}{c}{Year} & \multicolumn{1}{c}{Model Development} & \multicolumn{1}{c}{Map Assessment}\\
\midrule
2002 & 172 & \\
\addlinespace
2003 & 188 & \\
\addlinespace
2004 & 98 & \\
\addlinespace
2005 & 106 & \\
\addlinespace
2006 & 157 & \\
\addlinespace
2007 &  & 207\\
\addlinespace
2008 & 165 & \\
\addlinespace
2009 & 146 & \\
\addlinespace
2010 & 153 & \\
\addlinespace
2011 & 174 & \\
\addlinespace
2012 &  & 191\\
\addlinespace
2013 & 138 & \\
\addlinespace
2014 & 156 & \\
\addlinespace
2015 & 119 & \\
\addlinespace
2016 & 96 & \\
\addlinespace
2017 & 129 & \\
\addlinespace
2018 & 19 & 96\\
\addlinespace
2019 & 33 & 51\\
\midrule
\addlinespace
\textbf{Total} & \textbf{2049} & \textbf{545}\\
\bottomrule
\end{tabular}
\endgroup{}
\end{table}

\hypertarget{sec-lidar}{%
\subsection{LiDAR data and LiDAR pixel sampling}\label{sec-lidar}}

\begin{figure}

{\centering \includegraphics{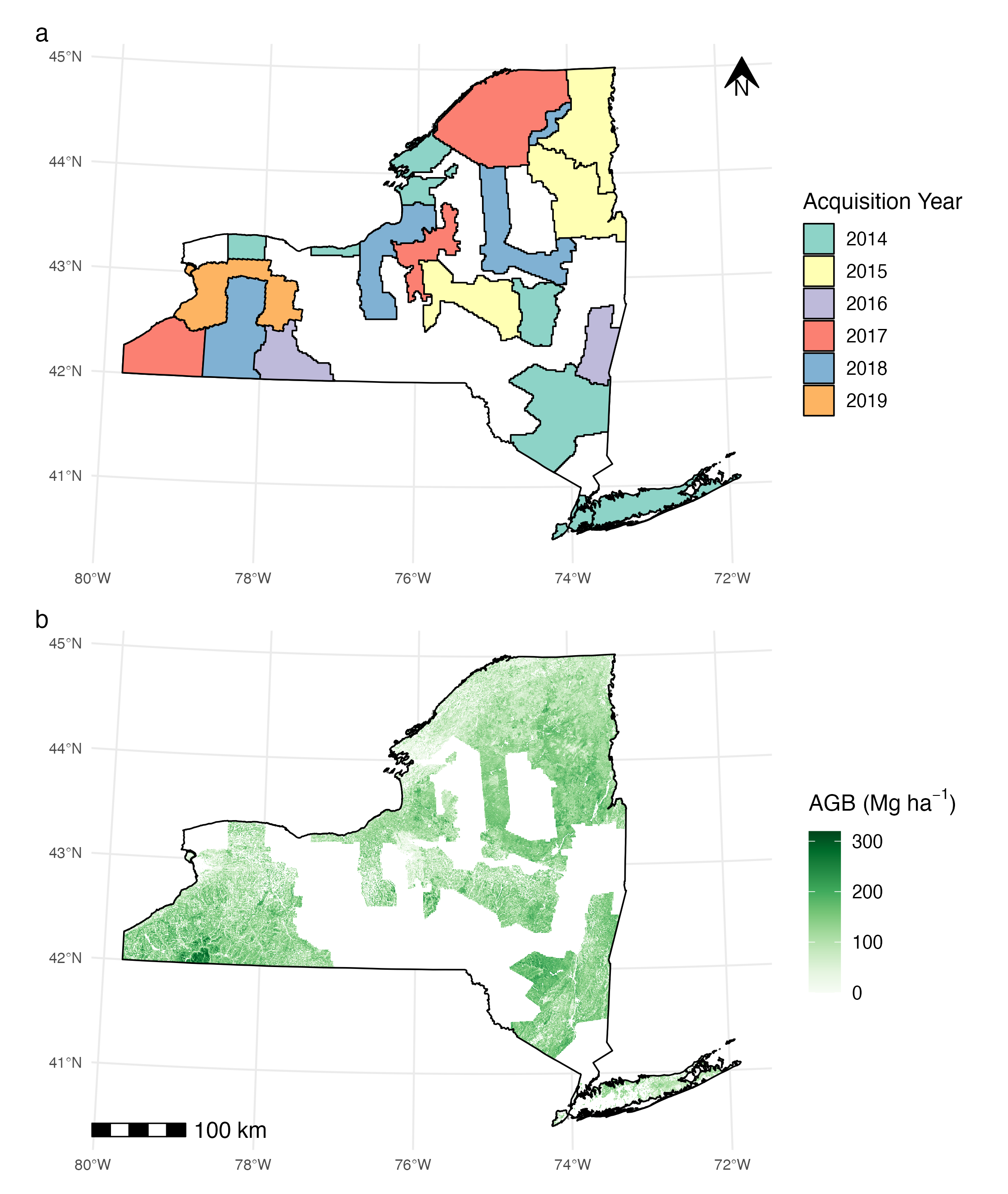}

}

\caption{\label{fig-lidarregions}LiDAR-based AGB reference data. a)
Spatial coverages of LiDAR collections colored by year of acquisition.
b) Spatiotemporal patchwork of LiDAR-based AGB predictions sampled for
reference data.}

\end{figure}

For our indirect modeling approach we used existing LiDAR-based AGB
prediction surfaces as reference data for model training
(Figure~\ref{fig-lidarregions}). Johnson et al. (2022) developed these
30 m surfaces with a spatio-temporal patchwork of 17 leaf-off LiDAR
collections covering 62.46\% (7,835,690 ha) of NYS. LiDAR data were
collected from altitudes ranging from 700--5300 m with pulse densities
ranging from 1.54--3.24 pulses per m\textsuperscript{2}. A set of 40
predictors computed from the height-normalized point clouds, in
combination with topographic, climatic, landcover, and cadastral data
were colocated with FIA plots as model training data. Stacked ensembles
(Wolpert 1992) of machine learning models were used to make predictions
across the patchwork; further details can be found in Johnson et al.
(2022).

Following Johnson et al. (2022), we restricted the map space using a
vegetation mask based on LCMAP primary classifications (Brown et al.
2020; Zhu and Woodcock 2014) as well as an area of applicability mask
(Meyer and Pebesma 2021). As such, our sample of LiDAR-based AGB
predictions was limited to vegetated landscapes, and where predictions
were based on predictor data that was sufficiently represented in the
training data. Following the indirect modeling efforts described in
Hudak et al. (2020), we conducted a stratified random sample from the
LiDAR-based AGB predictions, where strata were defined as 20 equal
intervals ranging from 0 to the maximum mapped AGB value
(\textasciitilde330 Mg ha\textsuperscript{-1}). 1,000 pixels were
sampled from each stratum resulting in a total of 20,000 spatially
resolved AGB predictions.

\hypertarget{sec-landsat-data}{%
\subsection{Landsat and auxiliary data}\label{sec-landsat-data}}

We produced a set of 16 annual Landsat-derived predictors by processing
Landsat collection 1 data (C1, USGS (2018)) in Google Earth Engine (GEE,
Gorelick et al. (2017)). We followed the processing framework described
in Michael J. Mahoney et al. (2022), relying on growing-season medoid
composites processed with coefficients from Roy et al. (2016) and the
Landtrendr implementation in GEE (hereafter LT-GEE) to provide a
continuous, and smoothed, 30-year time series of pixel-level metrics
describing surface conditions and disturbance history (Kennedy, Yang,
and Cohen 2010; Kennedy, Yang, et al. 2018). All spectral indices and
their respective deltas computed with a 1-year lag (Hudak et al. 2020)
were fit to Normalized Burn Ratio (NBR) temporally segmented vertices
(Kennedy, Yang, et al. 2018). We computed the normalized burn ratio
(NBR; Kauth and Thomas (1976)), tasseled-cap wetness, brightness, and
greenness (TCW, TCB, TCG; Cocke, Fulé, and Crouse (2005)), normalized
difference vegetation index (NDVI; Kriegler et al. (1969)), simple ratio
(SR; Jordan (1969)), and modified simple ratio (MSR; Chen (1996)) using
the `awesome-spectral-indices' javascript library for GEE (Montero et
al. 2022). The disturbance metrics were processed with a separate NBR
segmentation using LT-GEE parameters designed to be more sensitive to
the timing of discrete disturbance events (Kennedy, Yang, et al. 2018).
We chose to use NBR to process all other LT-GEE-derived predictors,
providing disturbance history and temporal break-points to which all
other indices were fit, since it has been demonstrated to best represent
disturbance events (Kennedy, Yang, and Cohen 2010). Supplementary
Materials 1 provides additional information on the LT-GEE parameters
used here.

We also included the annual primary and secondary land cover
classification predictions from United States Geological Survey's Land
Change Monitoring, Assessment, and Projection (LCMAP) version 1.2 (Brown
et al. 2020; Zhu and Woodcock 2014). Further, a set of steady-state
ancillary predictors was included to represent geospatial variation in
climate, topography, ecology, and landcover (Kennedy, Ohmann, et al.
2018). These predictors included precipitation and temperature 30 year
normals derived from PRISM Climate Group data (PRISM Climate Group
2022), elevation, aspect, slope, and a topographic wetness index derived
from a 30 m digital elevation model (Michael J. Mahoney, Beier, and
Ackerman 2022; U.S. Geological Survey 2019; Beven and Kirkby 1979), a
global canopy height map representing 2005 conditions (Simard et al.
2011; Hudak et al. 2020), distance (m) to nearest area and line water
identified by the US Census Bureau (Walker 2022; US Census Bureau 2013),
National Wetland Inventory classifications developed by the Fish and
Wildlife Services (FWS) (FWS 2022; Wilen and Bates 1995), and the
Environmental Protection Agency's (EPA) level 4 ecozones (Omernik and
Griffith 2014; CEC 1997). Where individual EPA level 4 ecozones did not
cover \(\geq\) 2\% of the state they were aggregated to their level 3
ecozone, and if this aggregation did not cover \(\geq\) 2\% of the state
these ecozones were set to ``other''. All categorical variables (LCMAP,
ecozones, wetlands) were encoded as boolean indicator variables.

Each of the 29 predictor layers (Table~\ref{tbl-preds}) were projected
to match Landsat 30 m pixel geometries. The raster stacks of predictors
were clipped and aggregated (weighted average) at the constructed FIA
plot polygons (Section~\ref{sec-fia}), and were also overlaid with the
sampled LiDAR-based AGB predictions (Section~\ref{sec-lidar}), creating
two distinct sets of data for model training based on the same set of
predictors. The exactextractr (Daniel Baston 2022) and terra (Hijmans
2022) packages for the R (R Core Team 2021) programming language were
used to compile the training datasets.

\hypertarget{tbl-preds}{}
\begin{table}[H]
\caption{\label{tbl-preds}Definitions of predictors used for model fitting. }\tabularnewline

\centering
\begin{tabular}[t]{>{\raggedright\arraybackslash}p{10em}>{\raggedright\arraybackslash}p{10em}>{\raggedright\arraybackslash}p{22em}}
\toprule
\multicolumn{1}{>{\centering\arraybackslash}p{10em}}{Group} & \multicolumn{1}{>{\centering\arraybackslash}p{10em}}{Predictor} & \multicolumn{1}{>{\centering\arraybackslash}p{22em}}{Definition}\\
\midrule
 & TCB, TCW, TCG & Tassled cap brightness, wetness, and greenness, with noise removed using LT-GEE\\
\cmidrule{2-3}
 & NBR & Normalized burn ratio with noise removed using LT-GEE\\
\cmidrule{2-3}
 & NDVI & Normalized difference vegetation index with noise removed using LT-GEE\\
\cmidrule{2-3}
 & SR & Simple ratio with noise removed using LT-GEE\\
\cmidrule{2-3}
\multirow[t]{-5}{10em}{\raggedright\arraybackslash Spectral indices} & MSR & Modified simple ratio with noise removed using LT-GEE\\
\cmidrule{1-3}
Delta & Delta\_* & Change computed with 1 year lag for all predictors in the 'Spectral indices' group\\
\cmidrule{1-3}
Disturbance & YOD, MAG & Year of most recent disturbance and associated magnitude of NBR change, as identified using an NBR segmentation in LT-GEE (1985-2019)\\
\cmidrule{1-3}
 & CHM & Global canopy height model reflecting 2005 conditions (Simard 2011), downsampled from 1 km to 30 m resolution\\
\cmidrule{2-3}
 & ECOZONE & EPA level 4 ecozones. Aggregated to level 3 if level 4 areas < 2\% of the state. Set to 'other' if level 3 aggregation < 2\% of state.\\
\cmidrule{2-3}
 & WETLAND & Wetland classification codes from the FWS National Wetlands Inventory\\
\cmidrule{2-3}
\multirow[t]{-4}{10em}{\raggedright\arraybackslash Ecological} & DIST\_TO\_WATER & Distance in meters to nearest TIGER/Line Shapefile water from the US Census Bureau\\
\cmidrule{1-3}
Climate & PRECIP, TMAX, TMIN & 30-year normals for precipitation, maximum temperature, and minimum temperature, derived from annual PRISM climate models\\
\cmidrule{1-3}
Topographic & ASPECT, ELEVATION, SLOPE, TWI & Aspect, elevation, slope, and topographic wetness index derived from a 30-meter digital elevation model\\
\cmidrule{1-3}
Landcover & LCPRI, LCSEC & LCMAP primary and secondary land cover classifications\\
\bottomrule
\end{tabular}
\end{table}

\hypertarget{sec-model-dev}{%
\subsection{Model development}\label{sec-model-dev}}

We developed three distinct modeling approaches using a standard
training framework. The direct approach involved training models on a
random 80\% partition of the model dataset derived from FIA field data
(Section~\ref{sec-fia}), and the indirect approach involved training
models on a random 80\% partition of the sample of LiDAR-based AGB
predictions (Section~\ref{sec-lidar}). We developed separate sets of ML
models for both approaches and combined each set in a stacked ensemble
to better reflect model selection uncertainty (Wintle et al. 2003) and
to reduce the generalization error of our component models (Wolpert
1992). The third approach was an ensemble combining predictions from the
direct and indirect ensemble models in a simple average, as model
averaging has been demonstrated to improve upon individual predictions
where data is noisy and the relationships between predictors and
responses are complex and largely unknown (Wolpert 1992; Dormann et al.
2018). For all three approaches, we used the 20\% test partitions to
assess model performance against each respective dataset and iterate
with various predictors and model forms.

Both the direct and indirect approaches used all 29 predictors described
in Section~\ref{sec-landsat-data}, while the ensemble was developed with
only predictions from these models. Both the direct and indirect
approaches combined a random forest, as implemented in the ranger R
package (Breiman 2001; Wright and Ziegler 2017) and a stochastic
gradient boosting machine (GBM) as implemented in the lightgbm R package
(Friedman 2002; Ke et al. 2017; Shi et al. 2022). The direct approach
also incorporated a support vector machine (SVM) as implemented in the
kernlab R package (Cortes and Vapnik 1995; Karatzoglou et al. 2004). SVM
training time scales between quadratic and cubic with respect to
training observations (Bottou and Lin 2007) and thus was not
computationally feasible to implement our indirect approach with 16,000
training points.

Each of the component ML models were tuned using the 80\% training
partition described above and an iterative grid search, starting by
testing wide ranges of hyperparameters using five-fold cross validation
and then narrowing down to only the most performant combinations over
several iterations. Models then used the most accurate sets of
hyperparameters in all other analyses. The selected hyperparameters for
each component model and the coefficients in the linear regression
ensembles are available in Supplementary Materials 2. For each of the n
observation in the training dataset, all component models were fit,
using their optimal hyperparameters, with n-1 observations. Predictions
for each component model were made for the nth (left out) observation. A
linear regression model was used to estimate AGB as a function of these
leave-one-out predictions, combining the component ML models in a linear
regression ensemble as follows:

\begin{equation}\protect\hypertarget{eq-ensemble}{}{
\operatorname{AGB} = 
\beta_{0} + \beta_{1} \cdot P_{1} + \ldots + \beta_{n} \cdot P_{n}
}\label{eq-ensemble}\end{equation}

where \(\beta_{*}\) are coefficients estimated through ordinary least
squares regression, and \(P_{*}\) are the respective component model
predictions. At an abstract level the direct approach was constructed as
follows:

\begin{equation}\protect\hypertarget{eq-direct}{}{
\operatorname{AGB} = ensemble(RF, SVM, LGB)
}\label{eq-direct}\end{equation}

where \(ensemble\) represents Equation~\ref{eq-ensemble}, and \(RF\),
\(SVM\), and \(GBM\) would be substituted for the \(P_{*}\) variables in
Equation~\ref{eq-ensemble}. The indirect approach was constructed as
follows: \begin{equation}\protect\hypertarget{eq-indirect}{}{
\operatorname{AGB} = ensemble(RF, LGB)
}\label{eq-indirect}\end{equation}

and the overarching ensemble was constructed as follows:

\begin{equation}\protect\hypertarget{eq-big-ensemble}{}{
\operatorname{AGB} = \frac{ensemble(RF_{direct}, SVM_{direct}, LGB_{direct}) + ensemble(RF_{indirect}, LGB_{indirect})}{2}
}\label{eq-big-ensemble}\end{equation}

\hypertarget{sec-maps}{%
\subsection{AGB mapping and postprocessing}\label{sec-maps}}

The linear model ensembles for the direct and indirect approaches, as
well as the overarching average ensemble, were used to make predictions
for all 30 m pixels across the state. With recognition that our
predictions are best suited to areas populated by woody biomass, we
overlaid our predictions with the LCMAP version 1.2 primary landcover
classification product (Brown et al. 2020; Zhu and Woodcock 2014), which
has a reported overall accuracy of 77.4\% in the Eastern United States
for the years 1985-2018 (Pengra et al. 2020). LCMAP data shared
identical pixel geometries with our AGB maps and its annual resolution
allowed for temporal alignment with each individual year of mapping. We
masked our AGB prediction surfaces to remove developed, cropland, water,
and barren pixels and then tabulated AGB by the three remaining
vegetated LCMAP classes of tree cover, grass/shrub, and wetland.

\hypertarget{sec-map-agree}{%
\subsection{Map agreement assessment}\label{sec-map-agree}}

We assessed the agreement between our AGB maps and FIA reference data
following approaches prescribed by Riemann et al. (2010) and Menlove and
Healey (2020). The former evaluated agreement across a range of scales
and accounts for the mismatch in spatial support between map aggregate
estimates (many pixels) and FIA aggregate estimates (few plots) by only
extracting pixels coincident with FIA plots. The latter compared
FIA-derived AGB estimates -- which have been adjusted for forest cover
within, and area-extrapolated to, hexagon map units -- to zonal averages
of our mapped AGB.

Following Riemann et al. (2010) we compared our AGB prediction surfaces
from each of the three modeling approaches to the map assessment dataset
(Section~\ref{sec-fia}). Comparisons were made at both the plot-to-pixel
scale and within variably-sized hexagons with distances between
centroids ranging from 20 km (34,641 ha) to 50 km (216,506 ha). Since
the plot inventories spanned multiple years (2007, 2012, 2018, 2019) we
extracted predictions from only those map surfaces that were temporally
aligned with the specific plot inventories in our dataset. We then
pooled this data together, producing a temporally generalized accuracy
assessment. As an extension of the Riemann et al. (2010) methodology we
assessed the spatial patterns of prediction error by summarizing the
plot-to-pixel residuals and FIA reference data distributions within
hexagon units with centroids spaced 50 km apart. We also grouped
plot-to-pixel results by the majority LCMAP classification at each plot,
to demonstrate the level of agreement across vegetated landcover
classes.

Following the Menlove and Healey (2020) approach, we compared the
average of our masked predictions, weighted by the proportion of each
pixel intersecting a given hexagon, to a set of FIA-derived estimates
for 64,000 ha hexagons representing FIA's finest acceptable scale for
the most recent inventory cycle in NYS (2013--2019). We used 2016 AGB
maps from each approach for this comparison since 2016 sits in the
center of the time period that is represented in the Menlove and Healey
(2020) data. As recommended, we accounted for differences in forest
definitions between the FIA estimates and our mapped estimates by
dividing FIA estimates by the proportion of vegetated (based on LCMAP
tree cover, grass/shrub, wetland) area within each hexagon. Lastly, we
limited this comparison to only hexagons with a majority area falling
inside NYS boundaries.

Assessment metrics included mean absolute error in Mg
ha\textsuperscript{-1} (MAE), percent MAE relative to mean reference AGB
(\% MAE), root-mean-squared error in Mg ha\textsuperscript{-1} (RMSE),
percent RMSE relative to mean reference AGB (\% RMSE), mean error in Mg
ha\textsuperscript{-1} (ME), and the coefficient of determination
(\(R^2\)). Equations and formulas for each metric and the associated
estimates of standard errors are provided in Supplementary Materials 3.
The exactextractr (Daniel Baston 2022), sf (Pebesma 2018), and terra
(Hijmans 2022) packages in the R programming language (R Core Team 2021)
were used to conduct all analyses described here.

\hypertarget{qualitative-comparisons-of-fine-spatial-patterns}{%
\subsection{Qualitative comparisons of fine spatial
patterns}\label{qualitative-comparisons-of-fine-spatial-patterns}}

We also visually compared mapped predictions for each modeling approach
in and around Huntington Wildlife Forest (HWF), a 6,000 ha forested area
in Newcomb, NYS containing both reserves and areas of active management
and where our team has developed a familiarity with the landscape
through in situ and remote observations alike. Though limited to a small
fraction of the statewide context, this comparison aimed to
qualitatively assess relative strengths and weaknesses in characterizing
fine spatial patterns of AGB density across various management regimes
and landscape conditions. We conducted pair-wise raster subtraction to
produce surfaces that highlighted areas of disagreement across modeling
approaches and used both a 1 m LiDAR-derived canopy height model (CHM;
Atlantic Inc (2015)) as well as 0.5 m natural color imagery from the
National Aerial Imagery Program (NAIP; Earth Resources Observation And
Science (EROS) Center (2017)) for additional qualitative reference
information. The CHM and the NAIP imagery reflected conditions in 2015,
and so 2015 AGB prediction surfaces from each modeling approach were
compared.

\hypertarget{results}{%
\section{Results}\label{results}}

\hypertarget{annual-aboveground-biomass-maps}{%
\subsection{Annual aboveground biomass
maps}\label{annual-aboveground-biomass-maps}}

\begin{figure}

{\centering \includegraphics{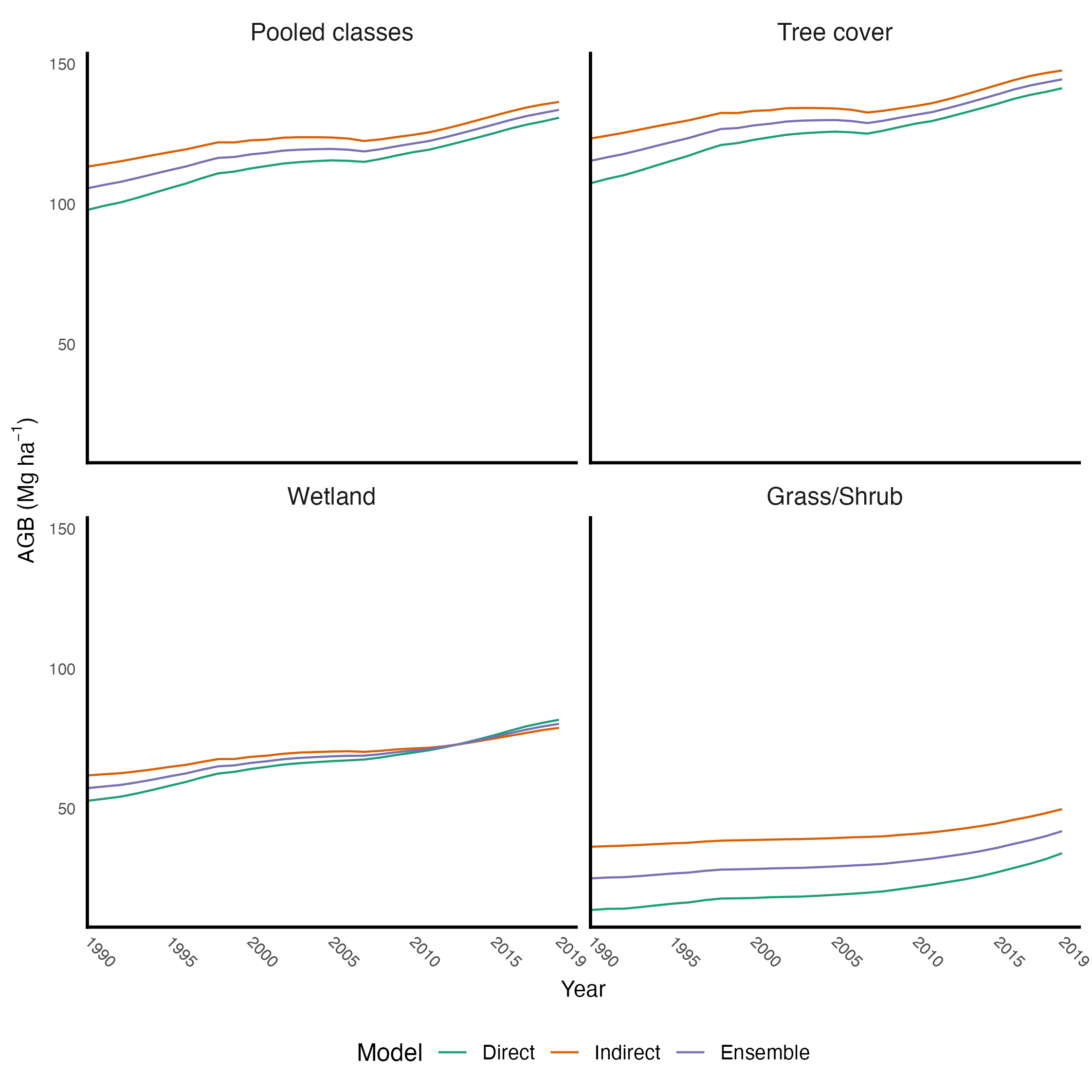}

}

\caption{\label{fig-ts}Annual statewide summaries (average AGB) for each
modeling approach by LCMAP class.}

\end{figure}

We produced 30 years (1990-2019) of statewide AGB maps at a 30 m
resolution using each of the three modeling approaches. Statewide AGB
averages for each of the three modeling approaches increased steadily
over the time period for each of the included LCMAP classifications
(Figure~\ref{fig-ts}). However, in agreement with its higher saturation
threshold (Section~\ref{sec-agree}), the indirect approach produced
significantly larger averages than both the direct and the ensemble
approaches (Figure~\ref{fig-ts}). Around 2006, all three models produced
small decreases in the statewide average for tree cover classified
pixels; this corresponds with the timing of large-scale insect outbreaks
in the Northeast (2005-2007, Kosiba et al. (2018)), and specifically a
forest tent caterpillar (\emph{Malacosoma disstria}) defoliation event
that affected roughly 1.2 million acres of land in NYS (USFS 2006).
While defoliation alone does not necessarily result in AGB loss, our
models' reliance on spectral information precluded them from making the
distinction between canopy-specific changes and structural changes.

\begin{figure}

{\centering \includegraphics{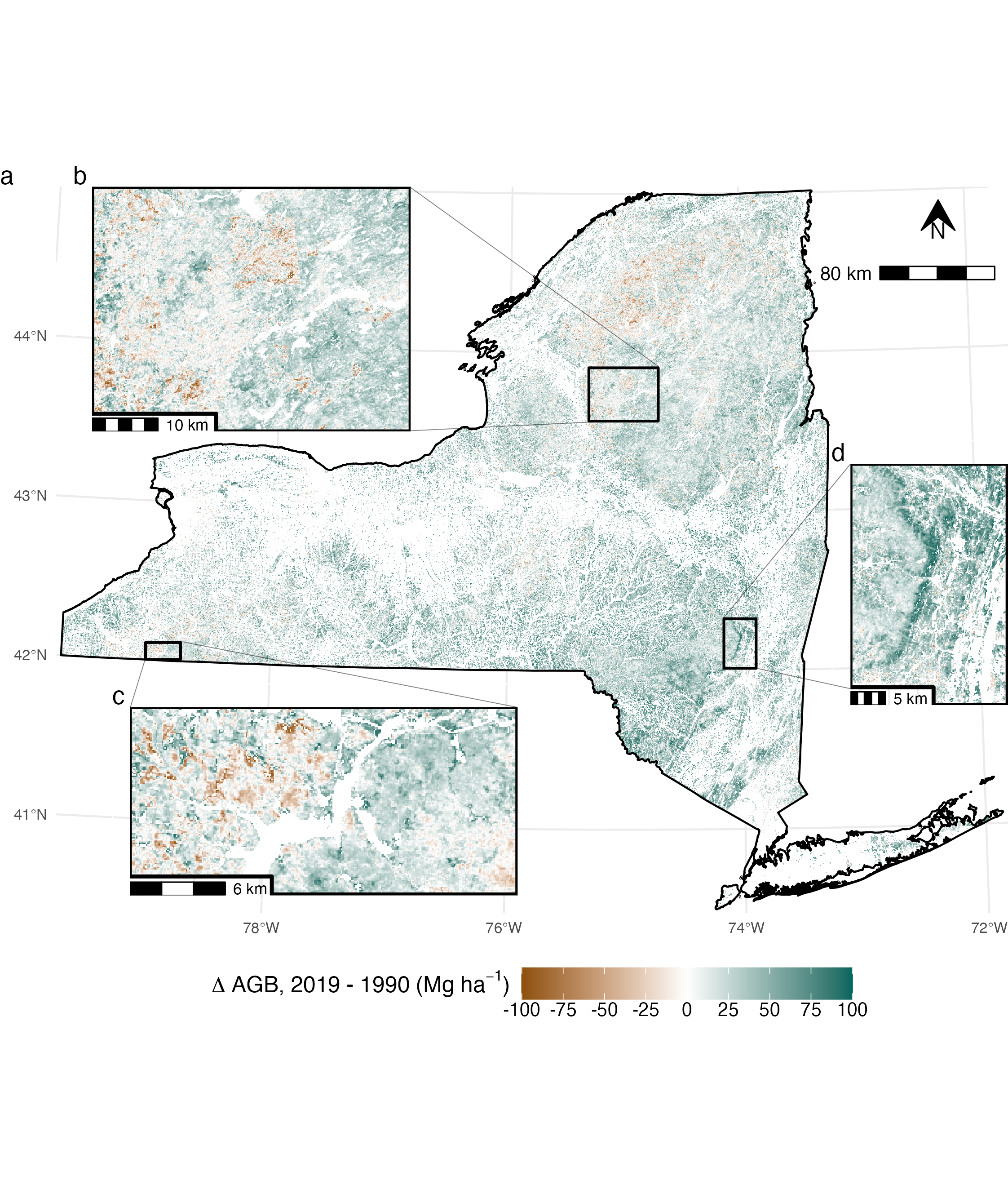}

}

\caption{\label{fig-state-agb-diff}New York State (USA) AGB difference
map (2019 AGB - 1990 AGB) with predictions from the ensemble model. a)
Statewide scale. b) A mosaic of working forests and Adirondack Forest
Preserve land south of Stillwater Reservoir, NYS. c) Allegany River area
with a portion of Allegany State Park to the east of the river. d)
Forest growth along the border of the Catskill Forest Preserve. Values
are capped at \(\pm\) 100 Mg ha\textsuperscript{-1} for display.}

\end{figure}

A full time timeseries raster subtraction (2019 AGB - 1990 AGB) using
the ensemble predictions reflected these annual trends, with increases
in AGB dominating the map (Figure~\ref{fig-state-agb-diff}). The 30-year
stock-change map also featured patterns of AGB change driven by
anthropogenic impacts and cadastral boundaries contrasted with those
that can be attributed to otherwise natural processes. Specifically, the
stock-change map highlighted a mosaic of working forests and Adirondack
Forest Preserve land and the varying spatial patterns and magnitudes of
change accompanying these distinct land uses
(Figure~\ref{fig-state-agb-diff} b), distinguished patchy AGB losses
within privately held lands to the west of the Allegany river against
subtle AGB gain and relative stability within Allegany State Park to the
east of the river (Figure~\ref{fig-state-agb-diff} c), and revealed a
band of forest growth that runs north to south along the border of the
Catskill Forest Preserve (Figure~\ref{fig-state-agb-diff} d).

\begin{figure}

{\centering \includegraphics{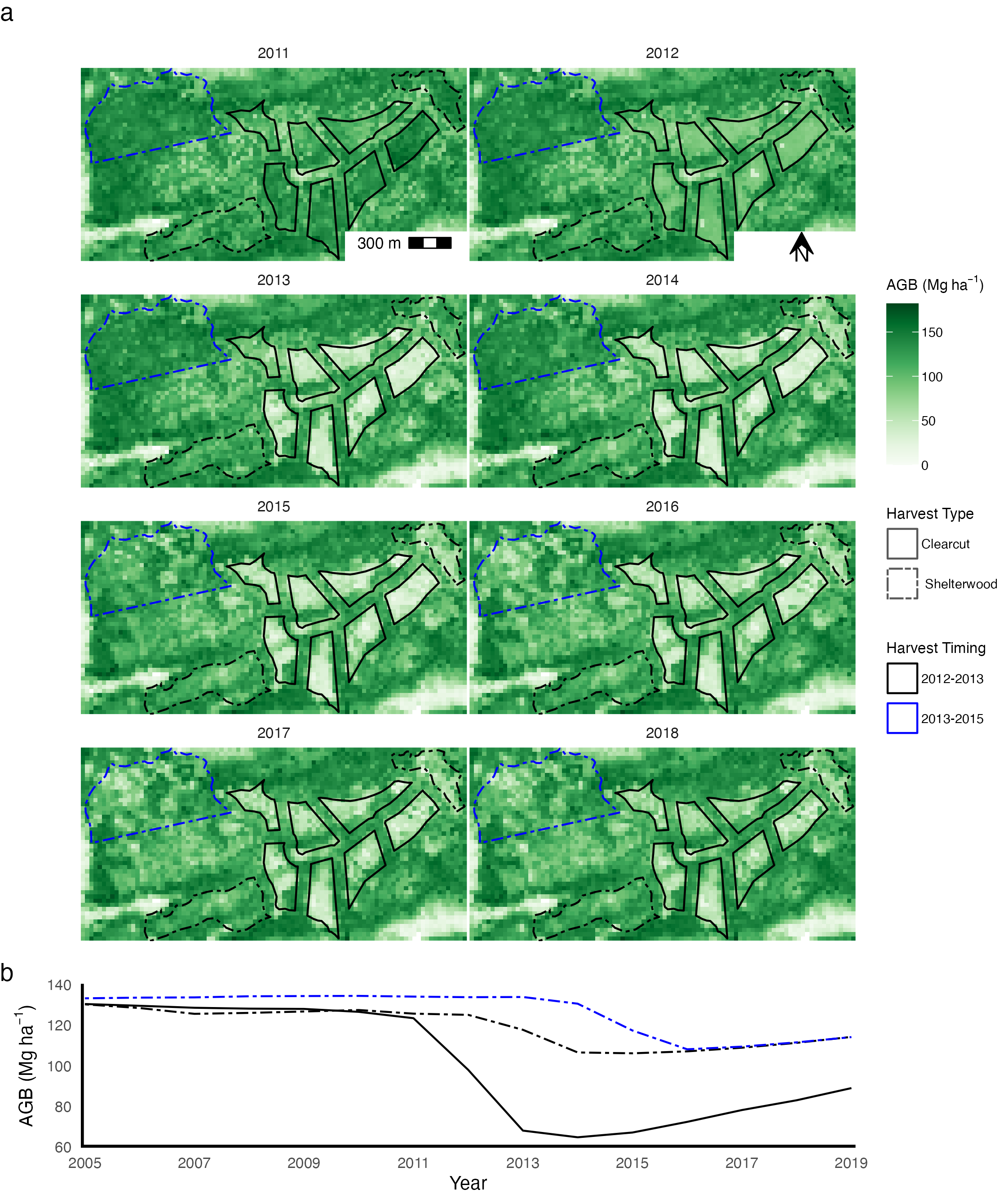}

}

\caption{\label{fig-harvest-agb}Quantifying AGB changes due to harvests
and subsequent regeneration in Northern New York State (USA). a) Annual
AGB predictions from the ensemble model for selected years overlaid with
harvest records symbolized by documented harvest type and timing. b)
Annual area-level summaries of mapped predictions (average AGB) for
harvest polygons grouped by harvest type and timing with trajectory
symbology corresponding to polygon symbology in a).}

\end{figure}

At the stand scale, where we have landowner-provided management records
in Northern NYS, our annual maps accurately captured the timing,
severity, and subsequent recovery (regeneration) from harvest activities
in working forests (Figure~\ref{fig-harvest-agb}). Looking in particular
at the clearcut harvests in Figure~\ref{fig-harvest-agb} and the
residual AGB within the boundaries of these polygons, we note that the
spatial management records we have are best approximations of harvest
prescriptions and may not reflect the true extent of harvest activity.
Likewise, disturbances outside these harvest polygons were captured in
our mapped predictions (note western portion of
Figure~\ref{fig-harvest-agb} beginning in 2015) and in this instance can
be attributed to harvest events that were simply not included in the
records provided by the landowner.

\hypertarget{sec-agree}{%
\subsection{Map agreement}\label{sec-agree}}

Although differences in estimated accuracy metrics were nominal among
our three modeling approaches, the ensemble model was most accurate
(Table~\ref{tbl-riemann}). The indirect approach on the other hand was
least accurate by these metrics, likely due to the additive effects of
pixel-level error in the initial LiDAR-AGB predictions (Johnson et al.
2022). We observed improved agreement between mapped predictions and FIA
estimates as the aggregation unit size increased for all three modeling
approaches, with \% MAE decreasing from 34.1 to 19.46\% for the direct
approach, from 35.69 to 20.23\% for the indirect approach, and from
33.88 to 19.23\% for the ensemble approach (Table~\ref{tbl-riemann}).
Similar patterns of increasing agreement were exhibited for MAE, RMSE,
\%RMSE, and R\textsuperscript{2}, but ME estimates were mostly stable
and positive across all scales of aggregation.

\begin{figure}

{\centering \includegraphics{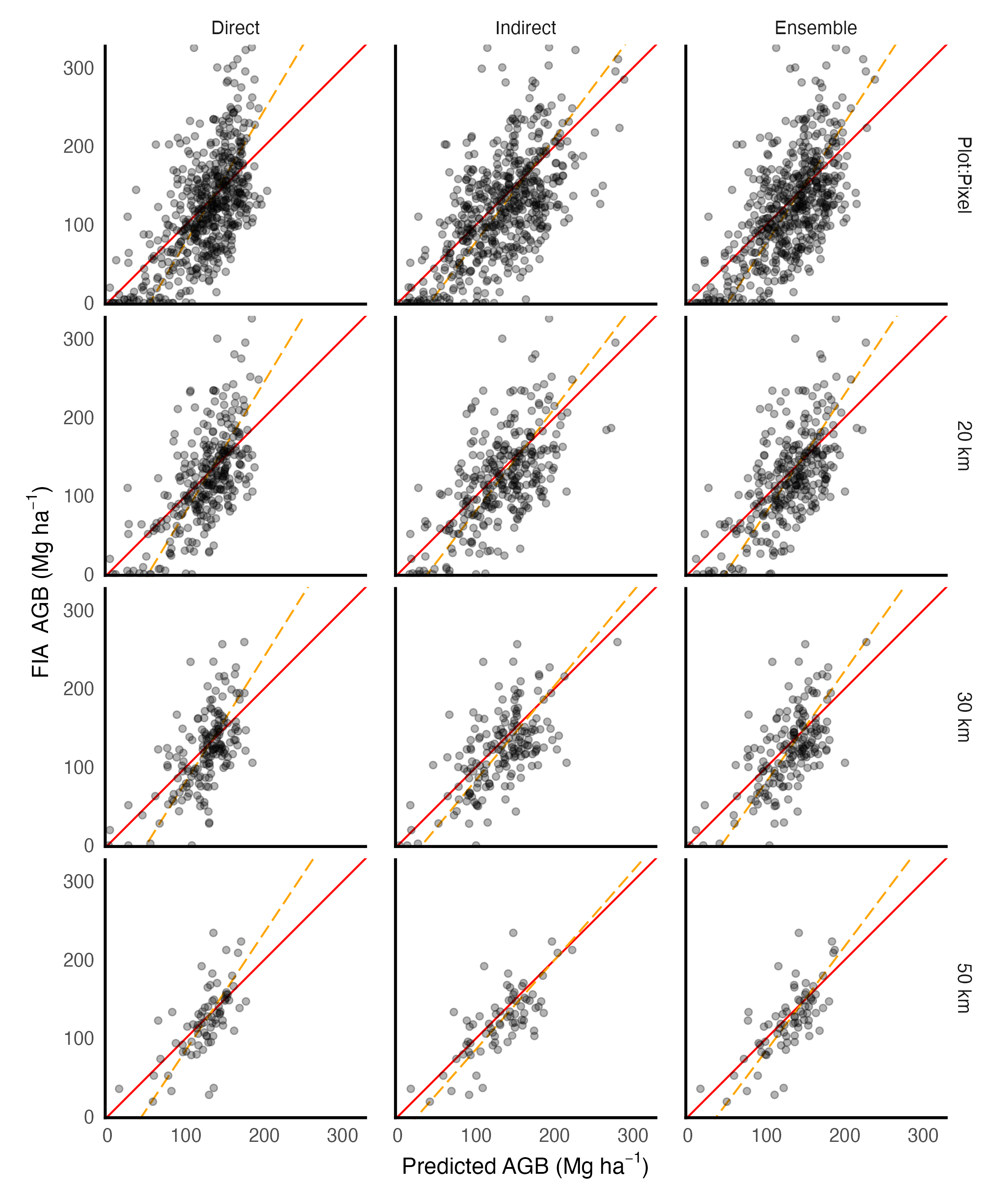}

}

\caption{\label{fig-riemann-11}Comparison of mapped AGB to FIA estimated
AGB across selected scales represented by distances between hexagon
centroids (plot:pixel, 10 km, 25 km, and 50 km). Geometric mean
functional relationship (GMFR) trend line shown with dashed (orange)
line, and 1:1 line shown with solid (red) line}

\end{figure}

All three models tended to overpredict on zero and near-zero AGB
reference observations, particularly at the plot:pixel and 20 km scales
of comparison (Figure~\ref{fig-riemann-11}), which resulted in positive
and significant ME estimates (Table~\ref{tbl-riemann}). Many of these
overpredictions can be explained by our reliance on tree-based models
(RF, GBM) whose predictions are the average values within terminal nodes
(Baccini et al. 2008; Urbazaev et al. 2018). However, these
overpredictions might also have been due to structural zeroes in our map
assessment dataset, where FIA AGB was assumed to be zero but is actually
not measured due to FIA's strict forest definition
(Section~\ref{sec-fia}; Johnson et al. (2022)). Large relative errors in
FIA plots classified as grass/shrub provided further evidence of the
impact of forest definition discrepancies on our map agreement results
(Table~\ref{tbl-lcmap-acc}). Unfortunately, we have had no means to
identify plots containing structural zeroes without additional data, and
could not separate them from plots with otherwise real overpredictions
and errors.

Underprediction on the largest reference observations (i.e.~saturation),
a common issue when modeling forest structure with optical imagery (Lu
2005; Duncanson, Niemann, and Wulder 2010), was evident for all three
modeling approaches but to varying degrees
(Figure~\ref{fig-riemann-11}). The direct approach saturated first,
failing to predict beyond 204 Mg ha\textsuperscript{-1}, whereas the
indirect approach was the best in this regard, predicting up to 289 Mg
ha\textsuperscript{-1} and leaving only 1\% of the reference data beyond
its ceiling. In general, patterns of over and underprediction diminished
and systematic agreement improved at larger scales of aggregation as
evidenced by the convergence of GMFR and 1:1 lines for all models
(Figure~\ref{fig-riemann-11}). The indirect approach yielded the best
systematic agreement (GMFR vs 1:1) across all scales despite being least
accurate in terms of the estimated metrics (Figure~\ref{fig-riemann-11};
Table~\ref{tbl-riemann}).

Map comparisons with the FIA's small area estimates (Menlove and Healey
(2020)) similarly demonstrated both patterns of over and under
prediction on the extremes of reference AGB distributions, as well as
the effects of saturation for each of the three modeling approaches
(Supplementary Materials 4). Despite consistently underpredicting
relative to the Menlove and Healey (2020) estimates, the direct approach
yielded more estimates within the provided 95\% confidence intervals
(90.31\%) as compared to the ensemble (88.27\%) and indirect (85.2\%)
approaches. Likewise, local errors (over and underprediction) were more
related to the amount of reference AGB within each hexagonal unit rather
than spatial or regional patterns when plot-to-pixel residuals were
mapped (Supplementary Materials 4).

Tree cover agreement for each model (Table~\ref{tbl-lcmap-acc}) largely
matched the overall plot-to-pixel agreement in Table~\ref{tbl-riemann},
because the vast majority of map assessment plots fell within this
classification. Map agreement was worse for the fewer number of wetland
and grass/shrub classified plots, with ME estimates indicating
significant overprediction in grass/shrub classified plots and
underprediction in wetland classified plots (Table~\ref{tbl-lcmap-acc}).
This discrepancy in agreement among vegetated classes can likely be
attributed to the varying degrees to which each landcover classification
was represented in our reference datasets and the mismatch between our
LCMAP-defined vegetation mask (Section~\ref{sec-maps}) and the strict
forest definition used by FIA (Section~\ref{sec-fia}).

\hypertarget{tbl-riemann}{}
\begin{table}
\caption{\label{tbl-riemann}Map agreement results for select scales. RMSE, MAE, ME in Mg ha\textsuperscript{-1}. Scale
= distance between hexagon centroids in km; PPH = plots per hexagon; n =
number of comparison units (plots or hexagons). All accuracy metrics as
defined in Supplementary Materials 3. Standard errors in parentheses
with minimum capped at 0.01. }\tabularnewline

\centering\begingroup\fontsize{10}{12}\selectfont

\resizebox{\linewidth}{!}{
\begin{tabular}[t]{lrrlrrrrrr}
\toprule
\multicolumn{1}{c}{Scale} & \multicolumn{1}{c}{n} & \multicolumn{1}{c}{PPH} & \multicolumn{1}{c}{Model} & \multicolumn{1}{c}{\% MAE} & \multicolumn{1}{c}{MAE} & \multicolumn{1}{c}{\% RMSE} & \multicolumn{1}{c}{RMSE} & \multicolumn{1}{c}{ME} & \multicolumn{1}{c}{R²}\\
\midrule
 &  &  & Direct & 34.10 & 41.20 (1.38) & 43.29 & 52.31 (3.06) & 4.83 (2.23) & 0.38 (0.01)\\

 &  &  & Indirect & 35.69 & 43.13 (1.44) & 45.30 & 54.73 (3.14) & 11.15 (2.30) & 0.32 (0.01)\\

\multirow{-3}{*}{\raggedright\arraybackslash Plot:Pixel} & \multirow{-3}{*}{\raggedleft\arraybackslash 545} & \multirow{-3}{*}{\raggedleft\arraybackslash } & Ensemble & 33.88 & 40.94 (1.36) & 42.84 & 51.76 (2.98) & 7.99 (2.19) & 0.39 (0.01)\\
\addlinespace
 &  &  & Direct & 29.17 & 35.53 (1.69) & 37.81 & 46.06 (4.19) & 3.42 (2.65) & 0.40 (0.01)\\

 &  &  & Indirect & 31.51 & 38.39 (1.71) & 39.80 & 48.48 (3.92) & 9.22 (2.74) & 0.33 (0.01)\\

\multirow{-3}{*}{\raggedright\arraybackslash 20 km} & \multirow{-3}{*}{\raggedleft\arraybackslash 302} & \multirow{-3}{*}{\raggedleft\arraybackslash 1.8} & Ensemble & 29.57 & 36.02 (1.64) & 37.65 & 45.86 (4.21) & 6.32 (2.62) & 0.40 (0.01)\\
\addlinespace
 &  &  & Direct & 25.35 & 30.76 (1.87) & 32.41 & 39.32 (6.12) & 3.66 (2.99) & 0.37 (0.01)\\

 &  &  & Indirect & 27.40 & 33.25 (1.86) & 33.97 & 41.21 (5.34) & 10.03 (3.06) & 0.30 (0.01)\\

\multirow{-3}{*}{\raggedright\arraybackslash 30 km} & \multirow{-3}{*}{\raggedleft\arraybackslash 172} & \multirow{-3}{*}{\raggedleft\arraybackslash 3.17} & Ensemble & 25.79 & 31.30 (1.76) & 32.02 & 38.86 (5.01) & 6.85 (2.93) & 0.38 (0.01)\\
\addlinespace
 &  &  & Direct & 19.46 & 23.85 (2.70) & 26.97 & 33.05 (12.80) & 2.58 (3.88) & 0.43 (0.01)\\

 &  &  & Indirect & 20.23 & 24.80 (2.39) & 26.14 & 32.04 (8.54) & 9.46 (3.61) & 0.46 (0.01)\\

\multirow{-3}{*}{\raggedright\arraybackslash 50 km} & \multirow{-3}{*}{\raggedleft\arraybackslash 73} & \multirow{-3}{*}{\raggedleft\arraybackslash 7.47} & Ensemble & 19.23 & 23.57 (2.39) & 25.38 & 31.10 (10.31) & 6.02 (3.60) & 0.49 (0.01)\\
\bottomrule
\end{tabular}}
\endgroup{}
\end{table}

\hypertarget{tbl-lcmap-acc}{}
\begin{table}
\caption{\label{tbl-lcmap-acc}Map agreement at the plot to pixel scale, grouped by LCMAP
classification. RMSE. MAE, ME in Mg ha\textsuperscript{-1}. n = number of plots. All
accuracy metrics as defined in Supplementary Materials 3. Standard
errors in parentheses with values capped at 0.01 and 1.00. }\tabularnewline

\centering
\resizebox{\linewidth}{!}{
\begin{tabular}[t]{lrlrrrrrr}
\toprule
\multicolumn{1}{c}{LCMAP} & \multicolumn{1}{c}{n} & \multicolumn{1}{c}{Model} & \multicolumn{1}{c}{\% MAE} & \multicolumn{1}{c}{MAE} & \multicolumn{1}{c}{\% RMSE} & \multicolumn{1}{c}{RMSE} & \multicolumn{1}{c}{ME} & \multicolumn{1}{c}{R²}\\
\midrule
 &  & Direct & 87.58 & 34.07 (7.50) & 111.81 & 43.49 (66.44) & 3.76 (12.02) & 0.41 (1.00)\\

 &  & Indirect & 101.20 & 39.36 (10.95) & 143.30 & 55.74 (217.09) & 33.85 (12.28) & 0.03 (1.00)\\

\multirow{-3}{*}{\raggedright\arraybackslash Grass/Shrub} & \multirow{-3}{*}{\raggedleft\arraybackslash 14} & Ensemble & 93.92 & 36.53 (6.65) & 112.34 & 43.69 (51.15) & 18.80 (10.94) & 0.40 (1.00)\\
\addlinespace
 &  & Direct & 40.19 & 34.56 (4.34) & 55.15 & 47.43 (28.20) & -9.21 (6.22) & 0.40 (0.01)\\

 &  & Indirect & 43.47 & 37.38 (4.26) & 57.12 & 49.12 (29.95) & -10.29 (6.42) & 0.35 (0.02)\\

\multirow{-3}{*}{\raggedright\arraybackslash Wetland} & \multirow{-3}{*}{\raggedleft\arraybackslash 57} & Ensemble & 40.87 & 35.15 (4.22) & 54.94 & 47.24 (31.20) & -9.75 (6.18) & 0.40 (0.01)\\
\addlinespace
 &  & Direct & 33.13 & 42.21 (1.48) & 41.67 & 53.10 (3.44) & 6.55 (2.42) & 0.31 (0.01)\\

 &  & Indirect & 34.47 & 43.93 (1.55) & 43.42 & 55.34 (3.46) & 13.06 (2.47) & 0.26 (0.01)\\

\multirow{-3}{*}{\raggedright\arraybackslash Tree cover} & \multirow{-3}{*}{\raggedleft\arraybackslash 474} & Ensemble & 32.78 & 41.77 (1.46) & 41.19 & 52.50 (3.23) & 9.80 (2.37) & 0.33 (0.01)\\
\bottomrule
\end{tabular}}
\end{table}

\hypertarget{qualitative-comparisons-of-fine-spatial-patterns-1}{%
\subsection{Qualitative comparisons of fine spatial
patterns}\label{qualitative-comparisons-of-fine-spatial-patterns-1}}

\begin{figure}

{\centering \includegraphics{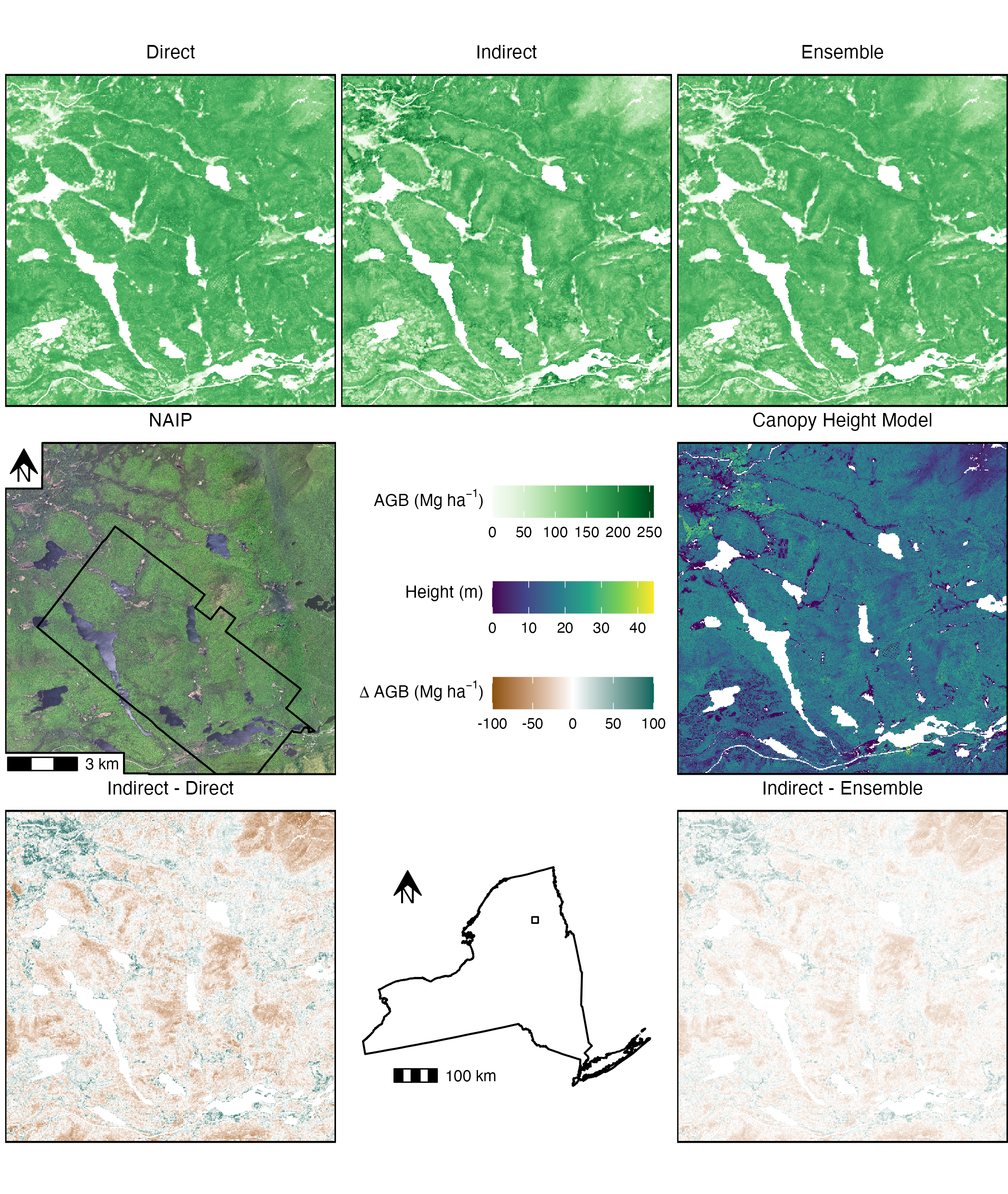}

}

\caption{\label{fig-hwf-compare}A qualitative comparison of maps from
each modeling approach within Huntington Wildlife Forest (boundary
mapped with black box in NAIP panel) and the surrounding area in
Newcomb, New York (full area extent mapped with black box in New York
State panel). Pair-wise raster subtractions (values capped at \(\pm\)
100 Mg ha\textsuperscript{-1} for display) highlight spatial patterns
and magnitudes of differences between model predictions. \emph{Ensemble
- Direct} not shown because it duplicates \emph{Indirect - Ensemble}. A
1 m LiDAR-derived canopy height model and 0.5 m natural color National
Aerial Imagery Program (NAIP) orthophotography included for additional
reference information. All surfaces represent conditions in 2015.}

\end{figure}

Within Huntington Wildlife Forest (HWF), in the forest preserve land to
the north of HWF (High Peaks Wilderness; Pataki and Cahill (1999)), and
in the working forest to the southwest of HWF, the indirect AGB map best
represented known patterns across the landscape and contained the most
spatial heterogeneity relative to the other two approaches
(Figure~\ref{fig-hwf-compare}). This was most evident where the largest
discrepancies between maps were present in the northeast and the
northwest corners of the area. In the northeast corner, where
conifer-dominated wetlands (NAIP Figure~\ref{fig-hwf-compare}) contained
some of the tallest vegetation in the area (CHM
Figure~\ref{fig-hwf-compare}), the indirect approach produced large
biomass predictions (\(\geq\) 225 Mg ha\textsuperscript{-1}) in
agreement with these landscape features. In the northwest corner of the
map, where high-elevation spruce-fir forests are present, the indirect
approach produced correspondingly small AGB predictions whereas the
direct approach was unable to distinguish these conditions from the rest
of the landscape. By definition, the ensemble map represented a blend of
characteristics from the direct and indirect maps in terms of both fine
spatial patterns and magnitudes of predictions.

\hypertarget{discussion}{%
\section{Discussion}\label{discussion}}

In this study we combined temporally smoothed, segmented, and gap-filled
Landsat imagery with a sample of LiDAR-based aboveground biomass (AGB)
predictions and a set of the USDA's Forest Inventory and Analysis (FIA)
field plots to produce annual wall-to-wall maps of AGB for New York
State (NYS), USA. To this end, we developed three separate modeling
approaches including direct, indirect, and ensemble approaches. Overall,
we found that all three modeling approaches performed similarly,
indicating that each approach could be satisfactory on its own, yet
tradeoffs were evident relating to model complexity, map accuracy,
saturation, and representation of fine spatial patterns. Comparisons to
existing studies with similar goals, but in temperate regions with
different disturbance and management regimes, indicated that the basic
methods herein can be leveraged to track forest biomass dynamics across
ecological domains and within working forests regardless of the
dominating forestry practices. The maps produced from each modeling
approach offer valuable insights into the spatiotemporal patterns of
forest structure, development, disturbance, and change over 30 years and
can serve as inputs for a variety of applications related to map-based
stock-change assessments, screening or prioritizing forest parcels for
enrollment in nature-based climate programs, and future monitoring,
reporting, and verification (MRV) systems across NYS.

\hypertarget{tradeoffs-among-modeling-approaches}{%
\subsection{Tradeoffs among modeling
approaches}\label{tradeoffs-among-modeling-approaches}}

There was no single winner among the three modeling approaches, but
rather each offered a set of benefits that can appeal to different
project-specific constraints and goals. Overall, the ensemble approach
produced the most accurate maps (Table~\ref{tbl-riemann}) which combined
characteristics from the direct and indirect approaches in terms of
fine-scale pattern representation and model saturation. Though it was
the most complex of the three approaches, it simultaneously mitigated
limitations and leveraged strengths associated with the plot-based
(Section~\ref{sec-fia}) and lidar-based (Section~\ref{sec-lidar})
training datasets. These results provide general support for model
ensembling in ecological applications where data are noisy and natural
variability is a significant source of error (Dormann et al. 2018).

By definition the direct approach was most parsimonious with only one
stage of modeling and the smallest investment of time and effort
required to produce AGB map products. The indirect and ensemble
approaches required the computationally demanding management and
analysis of terabytes of LiDAR data (Johnson et al. 2022), though that
effort could be reduced if LiDAR strips or samples were used in lieu of
wall to wall mapping (Wulder et al. 2012; Matasci et al. 2018; Urbazaev
et al. 2018). Additionally, increased complexity embedded in the
indirect and ensemble models makes estimating prediction uncertainty
more challenging than for the direct approach (Saarela et al. 2016).

The indirect approach was least impacted by saturation, resulting in the
best systematic agreement with FIA reference data across all scales
(Figure~\ref{fig-riemann-11}). With only 1\% of reference AGB plots
beyond the indirect model's prediction ceiling, this approach was best
suited to track continued growth in mature forest stands. This feature
would be especially important in NYS and the broader region where
historical land-use dynamics indicate that the majority of forest stands
have either reached or are approaching maturity
(Section~\ref{sec-study-area}). Failure to accurately quantify AGB in
these stands will lead to significant underestimation of carbon storage
and sequestration, at both local and statewide scales. Further, we found
that the indirect approach produced maps that best aligned with our
knowledge of local forest conditions and best represented fine-scaled
features on the landscape (Figure~\ref{fig-hwf-compare}). The strengths
of the indirect model can be attributed to the much larger sample of
reference data, and in theory the greater coverage of both the AGB
distribution and the landscape conditions in NYS, acquired from
broad-scale LiDAR-based AGB maps (Section~\ref{sec-lidar}).

\hypertarget{comparison-to-existing-studies}{%
\subsection{Comparison to existing
studies}\label{comparison-to-existing-studies}}

Comparisons of model performance and map agreement across studies should
be made with caution, as landscapes, data collection protocols, remotely
sensed data products, and AGB distributions can differ widely and have
large impacts on resulting agreement metrics. However, we do so here in
a relative fashion to situate the success of our approaches among
existing studies with similar goals. Kennedy, Ohmann, et al. (2018),
Hudak et al. (2020), and Matasci et al. (2018) each leveraged Landsat
time series data to map AGB annually at a 30 m resolution across the
following regions and time periods (respectively): Western Cascades
province of Oregon and Northern California, 2000-2016; Washington,
Oregon, Idaho, and Montana, 1990-2012; Canada's forest-dominated
ecosystems, 1984-2016. Kennedy, Ohmann, et al. (2018) used direct
modeling only, yielding an RMSE of \textasciitilde103 Mg
ha\textsuperscript{-1} against model training plots with a wide range of
AGB values (0-1000 Mg ha\textsuperscript{-1}), while Hudak et al. (2020)
and Matasci et al. (2018) exclusively used indirect modeling, yielding
64\% RMSE against independent FIA plots, and 66\% RMSE against
LiDAR-based AGB predictions respectively. Although these kinds of direct
comparisons have caveats, they signify that similar methods relying on
Landsat time series imagery to characterize forest dynamics are
applicable in multiple domains -- from conifer-dominated western US and
Canadian forests with even-aged disturbance regimes (Kennedy, Ohmann, et
al. 2018; White et al. 2017), to northern hardwoods and mixed forests of
the eastern US with mostly uneven-aged disturbance regimes
(Section~\ref{sec-study-area}). This capacity to track changes in
forests with varying disturbance patterns and management systems is
needed to ensure that all working forest landowners and landscapes are
treated accurately and fairly within large-scale carbon accounting
frameworks (Desrochers et al. 2022).

\hypertarget{applications-for-annual-agb-maps}{%
\subsection{Applications for annual AGB
maps}\label{applications-for-annual-agb-maps}}

Our rigorously evaluated map products have a range of applications where
knowledge of the spatiotemporal patterns of forest biomass (and by
extension, forest carbon pools) is needed. Most immediately, given our
extensive use of FIA plot-level information for model development
(Section~\ref{sec-fia}, Section~\ref{sec-lidar}) and map assessment
(Section~\ref{sec-map-agree}), our annual maps provide a translation of
FIA information to inputs for spatially explicit stock-change accounting
methods. Such a map-based framework offers the capability to summarize
stock changes and rates of sequestration following FIA's accounting
approach, but with the additional flexibility to do so for arbitrary
units of area within NYS for any time window in the 30 year period
(Figure~\ref{fig-state-agb-diff}). This increased resolution enabled the
identification of AGB losses and gains with distinct spatiotemporal
signatures attributed to conservation, regulation, and ownership
patterns across the landscape (Figure~\ref{fig-state-agb-diff} b, c, d).
While sample-based stock-change approaches will capture these outcomes
in aggregate, our maps can more precisely identify where, when, and how
both human and natural processes are impacting forest carbon stocks
across the landscape.

Although modeled data should not supersede direct measurements,
inventories, or boots-on-the-ground knowledge, the historical
perspective provided by our maps allows us to fill in gaps where
management records or forest inventory data are not available
(Figure~\ref{fig-harvest-agb}). The availability of both past management
information and historical AGB or carbon stock information opens the
door to a host of opportunities to quantify the outcomes of various
management regimes (Kaarakka et al. 2021; Patton et al. 2022). Further,
the burden of proving additionality for enrollment in carbon offset
programs hinges on establishing credible business-as-usual baselines
that are impossible to produce without historical data (Gillenwater et
al. 2007). Map datasets such as those developed here can fill this gap
for both potential enrollees and program managers alike, minimizing many
of the otherwise prohibitive up-front costs and requirements (Charnley,
Diaz, and Gosnell 2010; Kerchner and Keeton 2015). More broadly, these
historical datasets can provide baselines for better understanding
present and future forest conditions in response to multiple drivers of
change, including a rapidly changing climate (Cohen et al. 2016; White
et al. 2017).

Because we have primarily relied on federally funded and publicly
available data sources, as well as open source software and tools, we
have the flexibility to leverage the same methods developed for this
historical context to fulfill ongoing monitoring (MRV) needs. Our
modeling workflow needs only to be updated with annual Landsat imagery
and FIA inventories along with opportunistic additions of LiDAR
collections (Sugarbaker et al. 2014, 2017) to provide a highly
cost-effective landscape monitoring framework that is broadly
reproducible and extensible. This approach could be further enhanced by
integrating new streams of information that have the potential to
improve predictive accuracy relative to models trained with Landsat
alone (e.g.~ESA's Biomass mission -- Quegan et al. (2019), NASA's GEDI
mission -- Dubayah et al. (2014)).

Up-to-date maps of AGB and carbon stocks will allow decision-makers to
prioritize parcels for both protection via purchase of fee titles or
conservation easements, as well as for enrollment in improved forest
management programs and carbon markets (Merenlender et al. 2004;
Malmsheimer et al. 2008; Kelly, Germain, and Stehman 2015; Kerchner and
Keeton 2015). Similarly, timely annual AGB maps can support wall-to-wall
MRV and harvest monitoring, not necessarily in lieu of essential field
visits, but as a means to screen those parcels which are likely in
compliance from those which require a closer look (Gillenwater et al.
2007). Not only would monitoring costs be significantly reduced under
such a system, likely lowering financial break-even thresholds for
potential projects (Charnley, Diaz, and Gosnell 2010; Kerchner and
Keeton 2015), but strictly random site visits would also be rendered
dispensable when a regular census of properties or land holdings is
otherwise unfeasible. Beyond annual monitoring, fine-resolution AGB
trajectories derived from our 30 years of maps could inform timeseries
forecasting and landscape simulation studies that aim to predict the
carbon consequences of various policy and management scenarios (MacLean
et al. 2021).

\hypertarget{conclusion}{%
\section{Conclusion}\label{conclusion}}

Fine-resolution maps of historical forest dynamics can serve as inputs
to spatially explicit stock-change accounting frameworks that offer
critical information for projecting carbon outcomes of land stewardship
decisions at parcel to landscape scales. There is an essential need for
methods that can deliver these historical datasets in the near term and
that offer reproducible, consistent, and widely applicable data
products. We have demonstrated three model-based approaches leveraging
open source data, software, and tools to predict AGB annually, at a 30 m
resolution, across New York State (NYS) for the past three decades
(1990-2019). Our results show that each of the three approaches provide
valid outputs and offer unique benefits relative to each other, thus
offering a set of options for NYS where forests are expected to
contribute substantially as carbon sinks towards achieving a net-zero
carbon economy by 2050. More broadly, the map products produced here can
help managers and decision-makers maximize the role forested landscapes
will play in natural climate solutions and policies.

\hypertarget{acknowledgements}{%
\section{Acknowledgements}\label{acknowledgements}}

We would like to thank the USDA FIA program for their data sharing and
cooperation, the Dutch Pension Fund and F \& W Forest Management, LLC
for sharing management records, the NYS GPO for compiling and serving
LiDAR data, and the NYS Department of Environmental Conservation, Office
of Climate Change for funding support. We would also like to thank Grant
M. Domke, Stephen V. Stehman, and Eddie Bevilacqua for their support and
guidance throughout this project.

\newpage{}

\hypertarget{references}{%
\section*{References}\label{references}}
\addcontentsline{toc}{section}{References}

\hypertarget{refs}{}
\begin{CSLReferences}{1}{0}
\leavevmode\vadjust pre{\hypertarget{ref-ICESat-2}{}}%
Abdalati, Waleed, H. Jay Zwally, Robert Bindschadler, Bea Csatho, Sinead
Louise Farrell, Helen Amanda Fricker, David Harding, et al. 2010. {``The
{ICESat}-2 Laser Altimetry Mission.''} \emph{Proceedings of the {IEEE}}
98 (5): 735--51. \url{https://doi.org/10.1109/jproc.2009.2034765}.

\leavevmode\vadjust pre{\hypertarget{ref-wwe}{}}%
Atlantic Inc. 2015. {``{NY\_WarrenWashingtonEssex\_Spring2015}.''}
\url{ftp://ftp.gis.ny.gov/elevation/LIDAR/}.

\leavevmode\vadjust pre{\hypertarget{ref-Baccini2008}{}}%
Baccini, A, N Laporte, S J Goetz, M Sun, and H Dong. 2008. {``A First
Map of Tropical Africa's Above-Ground Biomass Derived from Satellite
Imagery.''} \emph{Environmental Research Letters} 3 (4): 045011.
\url{https://doi.org/10.1088/1748-9326/3/4/045011}.

\leavevmode\vadjust pre{\hypertarget{ref-Banskota2014}{}}%
Banskota, Asim, Nilam Kayastha, Michael J. Falkowski, Michael A. Wulder,
Robert E. Froese, and Joanne C. White. 2014. {``Forest Monitoring Using
Landsat Time Series Data: A Review.''} \emph{Canadian Journal of Remote
Sensing} 40 (5): 362--84.
\url{https://doi.org/10.1080/07038992.2014.987376}.

\leavevmode\vadjust pre{\hypertarget{ref-Bechtold2005}{}}%
Bechtold, William A, and Paul L Patterson. 2005. \emph{The Enhanced
Forest Inventory and Analysis Program--National Sampling Design and
Estimation Procedures}. Vol. 80. USDA Forest Service, Southern Research
Station. \url{https://doi.org/10.2737/SRS-GTR-80}.

\leavevmode\vadjust pre{\hypertarget{ref-Beven1979}{}}%
Beven, Keith J., and Mike J. Kirkby. 1979. {``A Physically Based,
Variable Contributing Area Model of Basin Hydrology.''}
\emph{Hydrological Sciences Bulletin} 24 (1): 43--69.
\url{https://doi.org/10.1080/02626667909491834}.

\leavevmode\vadjust pre{\hypertarget{ref-Bottou2007}{}}%
Bottou, Léon, and Chih-Jen Lin. 2007. {``Support Vector Machine
Solvers.''} \emph{Large Scale Kernel Machines} 3 (1): 301--20.
\url{https://web2.qatar.cmu.edu/~gdicaro/10315-Fall19/additional/SVM-solvers.pdf}.

\leavevmode\vadjust pre{\hypertarget{ref-Breiman2001}{}}%
Breiman, Leo. 2001. {``Random Forests.''} \emph{Machine Learning} 45
(1): 5--32. \url{https://doi.org/10.1023/A:1010933404324}.

\leavevmode\vadjust pre{\hypertarget{ref-Brown2020}{}}%
Brown, Jesslyn F., Heather J. Tollerud, Christopher P. Barber, Qiang
Zhou, John L. Dwyer, James E. Vogelmann, Thomas R. Loveland, et al.
2020. {``{Lessons learned implementing an operational continuous United
States national land change monitoring capability: The Land Change
Monitoring, Assessment, and Projection (LCMAP) approach}.''}
\emph{Remote Sensing of Environment} 238: 111356.
\url{https://doi.org/10.1016/j.rse.2019.111356}.

\leavevmode\vadjust pre{\hypertarget{ref-IPCC2019}{}}%
Buendia, E, K Tanabe, A Kranjc, J Baasansuren, M Fukuda, S Ngarize, A
Osako, Y Pyrozhenko, P Shermanau, and S Federici. 2019. {``Refinement to
the 2006 IPCC Guidelines for National Greenhouse Gas Inventories.''}
\emph{IPCC: Geneva, Switzerland} 5: 194.

\leavevmode\vadjust pre{\hypertarget{ref-CEC}{}}%
CEC. 1997. \emph{Ecological Regions of North America: Toward a Common
Perspective}. Commission for Environmental Cooperation (Montr{é}al,
Qu{é}bec).; Secretariat.

\leavevmode\vadjust pre{\hypertarget{ref-Charnley2010}{}}%
Charnley, Susan, David Diaz, and Hannah Gosnell. 2010. {``Mitigating
Climate Change Through Small-Scale Forestry in the {USA}: Opportunities
and Challenges.''} \emph{Small-Scale Forestry} 9 (4): 445--62.
\url{https://doi.org/10.1007/s11842-010-9135-x}.

\leavevmode\vadjust pre{\hypertarget{ref-Chen1996}{}}%
Chen, Jing M. 1996. {``Evaluation of Vegetation Indices and a Modified
Simple Ratio for Boreal Applications.''} \emph{Canadian Journal of
Remote Sensing} 22 (3): 229--42.
\url{https://doi.org/10.1080/07038992.1996.10855178}.

\leavevmode\vadjust pre{\hypertarget{ref-Cocke2005}{}}%
Cocke, Allison E., Peter Z. Fulé, and Joseph E. Crouse. 2005.
{``Comparison of Burn Severity Assessments Using Differenced Normalized
Burn Ratio and Ground Data.''} \emph{International Journal of Wildland
Fire} 14: 189--98. \url{https://doi.org/10.1071/WF04010}.

\leavevmode\vadjust pre{\hypertarget{ref-Cohen2016}{}}%
Cohen, Warren B., Zhiqiang Yang, Stephen V. Stehman, Todd A. Schroeder,
David M. Bell, Jeffrey G. Masek, Chengquan Huang, and Garrett W. Meigs.
2016. {``Forest Disturbance Across the Conterminous United States from
1985{\textendash}2012: The Emerging Dominance of Forest Decline.''}
\emph{Forest Ecology and Management} 360 (January): 242--52.
\url{https://doi.org/10.1016/j.foreco.2015.10.042}.

\leavevmode\vadjust pre{\hypertarget{ref-Cooke2000}{}}%
Cooke, William H. 2000. {``Forest/Non-Forest Stratification in Georgia
with Landsat Thematic Mapper Data.''} In \emph{McRoberts, Ronald e.;
Reams, Gregory a.; van Deusen, Paul c., Eds. Proceedings of the First
Annual Forest Inventory and Analysis Symposium; Gen. Tech. Rep. NC-213.
St. Paul, MN: US Department of Agriculture, Forest Service, North
Central Research Station: 28-30}.

\leavevmode\vadjust pre{\hypertarget{ref-Cortes1995}{}}%
Cortes, Corinna, and Vladimir Vapnik. 1995. {``Support-Vector
Networks.''} \emph{Machine Learning} 20 (3): 273--97.
\url{https://doi.org/10.1007/bf00994018}.

\leavevmode\vadjust pre{\hypertarget{ref-exactextractr}{}}%
Daniel Baston. 2022. \emph{Exactextractr: Fast Extraction from Raster
Datasets Using Polygons}.
\url{https://CRAN.R-project.org/package=exactextractr}.

\leavevmode\vadjust pre{\hypertarget{ref-Desrochers2022}{}}%
Desrochers, Madeleine L, Wayne Tripp, Stephen Logan, Eddie Bevilacqua,
Lucas Johnson, and Colin M Beier. 2022. {``Ground-Truthing Forest Change
Detection Algorithms in Working Forests of the {US} Northeast.''}
\emph{Journal of Forestry} 120 (5): 575--87.
\url{https://doi.org/10.1093/jofore/fvab075}.

\leavevmode\vadjust pre{\hypertarget{ref-Dormann2018}{}}%
Dormann, Carsten F., Justin M. Calabrese, Gurutzeta Guillera-Arroita,
Eleni Matechou, Volker Bahn, Kamil Bartoń, Colin M. Beale, et al. 2018.
{``Model Averaging in Ecology: A Review of Bayesian,
Information-Theoretic, and Tactical Approaches for Predictive
Inference.''} \emph{Ecological Monographs} 88 (4): 485--504.
\url{https://doi.org/10.1002/ecm.1309}.

\leavevmode\vadjust pre{\hypertarget{ref-GEDI}{}}%
Dubayah, Ralph, Scott J Goetz, James Bryan Blair, TE Fatoyinbo, Matthew
Hansen, Sean P Healey, Michelle A Hofton, et al. 2014. {``The Global
Ecosystem Dynamics Investigation.''} In \emph{AGU Fall Meeting
Abstracts}, 2014:U14A--07.

\leavevmode\vadjust pre{\hypertarget{ref-Duncanson2010}{}}%
Duncanson, LI, KO Niemann, and MA Wulder. 2010. {``Integration of GLAS
and Landsat TM Data for Aboveground Biomass Estimation.''}
\emph{Canadian Journal of Remote Sensing} 36 (2): 129--41.
\url{https://doi.org/10.5589/m10-037}.

\leavevmode\vadjust pre{\hypertarget{ref-Dyer2006}{}}%
Dyer, James M. 2006. {``Revisiting the Deciduous Forests of Eastern
North America.''} \emph{BioScience} 56 (4): 341--52.
\url{https://doi.org/10.1641/0006-3568(2006)56\%5B341:RTDFOE\%5D2.0.CO;2}.

\leavevmode\vadjust pre{\hypertarget{ref-NAIP}{}}%
Earth Resources Observation And Science (EROS) Center. 2017. {``National
Agriculture Imagery Program (NAIP).''} U.S. Geological Survey.
\url{https://doi.org/10.5066/F7QN651G}.

\leavevmode\vadjust pre{\hypertarget{ref-Fargione2018}{}}%
Fargione, Joseph E., Steven Bassett, Timothy Boucher, Scott D. Bridgham,
Richard T. Conant, Susan C. Cook-Patton, Peter W. Ellis, et al. 2018.
{``Natural Climate Solutions for the United States.''} \emph{Science
Advances} 4 (11). \url{https://doi.org/10.1126/sciadv.aat1869}.

\leavevmode\vadjust pre{\hypertarget{ref-Friedman2002}{}}%
Friedman, Jerome H. 2002. {``Stochastic Gradient Boosting.''}
\emph{Computational Statistics and Data Analysis} 38 (4): 367--78.
\url{https://doi.org/10.1016/S0167-9473(01)00065-2}.

\leavevmode\vadjust pre{\hypertarget{ref-NWI}{}}%
FWS. 2022. \emph{National Wetlands Inventory}. U.S. Department of the
Interior, Fish; Wildlife Service, Washington, D.C.
\url{https://fwsprimary.wim.usgs.gov/wetlands/apps/wetlands-mapper/}.

\leavevmode\vadjust pre{\hypertarget{ref-Gillenwater2007}{}}%
Gillenwater, Michael, Derik Broekhoff, Mark Trexler, Jasmine Hyman, and
Rob Fowler. 2007. {``Policing the Voluntary Carbon Market.''}
\emph{Nature Climate Change} 1 (711): 85--87.
\url{https://doi.org/10.1038/climate.2007.58}.

\leavevmode\vadjust pre{\hypertarget{ref-Gorelick2017}{}}%
Gorelick, Noel, Matt Hancher, Mike Dixon, Simon Ilyushchenko, David
Thau, and Rebecca Moore. 2017. {``Google Earth Engine: Planetary-Scale
Geospatial Analysis for Everyone.''} \emph{Remote Sensing of
Environment} 202: 18--27.
\url{https://doi.org/10.1016/j.rse.2017.06.031}.

\leavevmode\vadjust pre{\hypertarget{ref-Gray2012}{}}%
Gray, Andrew N, Thomas J Brandeis, John D Shaw, William H McWilliams,
and Patrick Miles. 2012. {``{Forest Inventory and Analysis Database of
the United States of America (FIA)}.''} \emph{Biodiversity and Ecology}
4: 225--31. \url{https://doi.org/10.7809/b-e.00079}.

\leavevmode\vadjust pre{\hypertarget{ref-Hansen2012}{}}%
Hansen, Matthew C., and Thomas R. Loveland. 2012. {``A Review of Large
Area Monitoring of Land Cover Change Using Landsat Data.''} \emph{Remote
Sensing of Environment} 122 (July): 66--74.
\url{https://doi.org/10.1016/j.rse.2011.08.024}.

\leavevmode\vadjust pre{\hypertarget{ref-Heath2009}{}}%
Heath, Linda S, Mark Hansen, James E Smith, Patrick D Miles, and Brad W
Smith. 2009. {``Investigation into Calculating Tree Biomass and Carbon
in the FIADB Using a Biomass Expansion Factor Approach.''} In
\emph{Proceedings of the Forest Inventory and Analysis (FIA) Symposium},
21--23.

\leavevmode\vadjust pre{\hypertarget{ref-terra}{}}%
Hijmans, Robert J. 2022. \emph{Terra: Spatial Data Analysis}.
\url{https://CRAN.R-project.org/package=terra}.

\leavevmode\vadjust pre{\hypertarget{ref-Hoppus2005}{}}%
Hoppus, Michael, and Andrew Lister. 2005. {``The Status of Accurately
Locating Forest Inventory and Analysis Plots Using the Global
Positioning System.''} In \emph{Proceedings of the Seventh Annual Forest
Inventory and Analysis Symposium}.
\url{https://www.nrs.fs.fed.us/pubs/7040}.

\leavevmode\vadjust pre{\hypertarget{ref-Houghton2005}{}}%
Houghton, R. A. 2005. {``Aboveground Forest Biomass and the Global
Carbon Balance.''} \emph{Global Change Biology} 11 (6): 945--58.
\url{https://doi.org/10.1111/j.1365-2486.2005.00955.x}.

\leavevmode\vadjust pre{\hypertarget{ref-Houghton2009}{}}%
Houghton, R. A., Forrest Hall, and Scott J. Goetz. 2009. {``Importance
of Biomass in the Global Carbon Cycle.''} \emph{Journal of Geophysical
Research: Biogeosciences} 114 (G2): n/a--.
\url{https://doi.org/10.1029/2009jg000935}.

\leavevmode\vadjust pre{\hypertarget{ref-Houghton2012}{}}%
Houghton, R. A., J. I. House, J. Pongratz, G. R. van der Werf, R. S.
DeFries, M. C. Hansen, C. Le Quéré, and N. Ramankutty. 2012. {``Carbon
Emissions from Land Use and Land-Cover Change.''} \emph{Biogeosciences}
9 (12): 5125--42. \url{https://doi.org/10.5194/bg-9-5125-2012}.

\leavevmode\vadjust pre{\hypertarget{ref-Huang2019}{}}%
Huang, Wenli, Katelyn Dolan, Anu Swatantran, Kristofer Johnson, Hao
Tang, Jarlath O'Neil-Dunne, Ralph Dubayah, and George Hurtt. 2019.
{``{High-resolution mapping of aboveground biomass for forest carbon
monitoring system in the Tri-State region of Maryland, Pennsylvania and
Delaware, {USA}}.''} \emph{Environmental Research Letters} 14 (9):
095002. \url{https://doi.org/10.1088/1748-9326/ab2917}.

\leavevmode\vadjust pre{\hypertarget{ref-Hudak2020}{}}%
Hudak, Andrew T, Patrick A Fekety, Van R Kane, Robert E Kennedy, Steven
K Filippelli, Michael J Falkowski, Wade T Tinkham, et al. 2020. {``A
Carbon Monitoring System for Mapping Regional, Annual Aboveground
Biomass Across the Northwestern {USA}.''} \emph{Environmental Research
Letters} 15 (9): 095003. \url{https://doi.org/10.1088/1748-9326/ab93f9}.

\leavevmode\vadjust pre{\hypertarget{ref-Johnson2022}{}}%
Johnson, Lucas K., Michael J. Mahoney, Eddie Bevilacqua, Stephen V.
Stehman, Grant M. Domke, and Colin M. Beier. 2022. {``Fine-Resolution
Landscape-Scale Biomass Mapping Using a Spatiotemporal Patchwork of
LiDAR Coverages.''} \emph{International Journal of Applied Earth
Observation and Geoinformation} 114: 103059.
\url{https://doi.org/10.1016/j.jag.2022.103059}.

\leavevmode\vadjust pre{\hypertarget{ref-Jordan1969}{}}%
Jordan, Carl F. 1969. {``Derivation of Leaf-Area Index from Quality of
Light on the Forest Floor.''} \emph{Ecology} 50 (4): 663--66.
\url{https://doi.org/10.2307/1936256}.

\leavevmode\vadjust pre{\hypertarget{ref-Kaarakka2021}{}}%
Kaarakka, Lilli, Meredith Cornett, Grant Domke, Todd Ontl, and Laura E.
Dee. 2021. {``Improved Forest Management as a Natural Climate Solution:
A Review.''} \emph{Ecological Solutions and Evidence} 2 (3).
\url{https://doi.org/10.1002/2688-8319.12090}.

\leavevmode\vadjust pre{\hypertarget{ref-kernlab}{}}%
Karatzoglou, Alexandros, Alex Smola, Kurt Hornik, and Achim Zeileis.
2004. {``Kernlab -- an {S4} Package for Kernel Methods in {R}.''}
\emph{Journal of Statistical Software} 11 (9): 1--20.
\url{https://doi.org/10.18637/jss.v011.i09}.

\leavevmode\vadjust pre{\hypertarget{ref-Kauth1976}{}}%
Kauth, Richard J., and G. S. P. Thomas. 1976. {``The Tasselled Cap - a
Graphic Description of the Spectral-Temporal Development of Agricultural
Crops as Seen by Landsat.''} In \emph{Symposium on Machine Processing of
Remotely Sensed Data}.

\leavevmode\vadjust pre{\hypertarget{ref-Guolin2017}{}}%
Ke, Guolin, Qi Meng, Thomas Finley, Taifeng Wang, Wei Chen, Weidong Ma,
Qiwei Ye, and Tie-Yan Liu. 2017. {``{LightGBM: A Highly Efficient
Gradient Boosting Decision Tree}.''} In \emph{Advances in Neural
Information Processing Systems}, edited by I. Guyon, U. V. Luxburg, S.
Bengio, H. Wallach, R. Fergus, S. Vishwanathan, and R. Garnett. Vol. 30.
Curran Associates, Inc.
\url{https://proceedings.neurips.cc/paper/2017/file/6449f44a102fde848669bdd9eb6b76fa-Paper.pdf}.

\leavevmode\vadjust pre{\hypertarget{ref-Kelly2015}{}}%
Kelly, Matthew C, René H Germain, and Stephen V Stehman. 2015. {``Family
Forest Owner Preferences for Forest Conservation Programs: A New York
Case Study.''} \emph{Forest Science} 61 (3): 597--603.
\url{https://doi.org/10.5849/forsci.13-120}.

\leavevmode\vadjust pre{\hypertarget{ref-Kennedy2018b}{}}%
Kennedy, Robert E, Janet Ohmann, Matt Gregory, Heather Roberts, Zhiqiang
Yang, David M Bell, Van Kane, et al. 2018. {``An Empirical, Integrated
Forest Biomass Monitoring System.''} \emph{Environmental Research
Letters} 13 (2): 025004. \url{https://doi.org/10.1088/1748-9326/aa9d9e}.

\leavevmode\vadjust pre{\hypertarget{ref-Kennedy2010}{}}%
Kennedy, Robert E, Zhiqiang Yang, and Warren B. Cohen. 2010.
{``Detecting Trends in Forest Disturbance and Recovery Using Yearly
Landsat Time Series: 1. {LandTrendr} {\textemdash} Temporal Segmentation
Algorithms.''} \emph{Remote Sensing of Environment} 114 (12):
2897--2910. \url{https://doi.org/10.1016/j.rse.2010.07.008}.

\leavevmode\vadjust pre{\hypertarget{ref-Kennedy2018}{}}%
Kennedy, Robert E, Zhiqiang Yang, Noel Gorelick, Justin Braaten, Lucas
Cavalcante, Warren B. Cohen, and Sean Healey. 2018. {``Implementation of
the LandTrendr Algorithm on Google Earth Engine.''} \emph{Remote
Sensing} 10 (5). \url{https://doi.org/10.3390/rs10050691}.

\leavevmode\vadjust pre{\hypertarget{ref-Kerchner2015}{}}%
Kerchner, Charles D., and William S. Keeton. 2015.
{``California{\textquotesingle}s Regulatory Forest Carbon Market:
Viability for Northeast Landowners.''} \emph{Forest Policy and
Economics} 50 (January): 70--81.
\url{https://doi.org/10.1016/j.forpol.2014.09.005}.

\leavevmode\vadjust pre{\hypertarget{ref-Kosiba2018}{}}%
Kosiba, Alexandra M., Garrett W. Meigs, James A. Duncan, Jennifer A.
Pontius, William S. Keeton, and Emma R. Tait. 2018. {``Spatiotemporal
Patterns of Forest Damage and Disturbance in the Northeastern United
States: 2000{\textendash}2016.''} \emph{Forest Ecology and Management}
430 (December): 94--104.
\url{https://doi.org/10.1016/j.foreco.2018.07.047}.

\leavevmode\vadjust pre{\hypertarget{ref-Kriegler1969}{}}%
Kriegler, Frank J., William A. Malila, Richard F. Nalepka, and W.
Richardson. 1969. {``Preprocessing Transformations and Their Effects on
Multispectral Recognition.''} In.

\leavevmode\vadjust pre{\hypertarget{ref-Lorimer2001}{}}%
Lorimer, Craig G. 2001. {``Historical and Ecological Roles of
Disturbance in Eastern North American Forests: 9,000 Years of Change.''}
\emph{Wildlife Society Bulletin (1973-2006)} 29 (2): 425--39.
\url{http://www.jstor.org/stable/3784167}.

\leavevmode\vadjust pre{\hypertarget{ref-Lu2005}{}}%
Lu, Dengsheng. 2005. {``Aboveground Biomass Estimation Using Landsat TM
Data in the Brazilian Amazon.''} \emph{International Journal of Remote
Sensing} 26 (12): 2509--25.
\url{https://doi.org/10.1080/01431160500142145}.

\leavevmode\vadjust pre{\hypertarget{ref-GrahamMacLean2021}{}}%
MacLean, Meghan Graham, Matthew J. Duveneck, Joshua Plisinski, Luca L.
Morreale, Danelle Laflower, and Jonathan R. Thompson. 2021. {``Forest
Carbon Trajectories: Consequences of Alternative Land-Use Scenarios in
New England.''} \emph{Global Environmental Change} 69 (July): 102310.
\url{https://doi.org/10.1016/j.gloenvcha.2021.102310}.

\leavevmode\vadjust pre{\hypertarget{ref-terrainr}{}}%
Mahoney, Michael J., Colin M. Beier, and Aidan C. Ackerman. 2022.
{``{terrainr}: An {R} Package for Creating Immersive Virtual
Environments.''} \emph{Journal of Open Source Software} 7 (69): 4060.
\url{https://doi.org/10.21105/joss.04060}.

\leavevmode\vadjust pre{\hypertarget{ref-Mahoney2022}{}}%
Mahoney, Michael J, Lucas K Johnson, Abigail Z Guinan, and Colin M
Beier. 2022. {``Classification and Mapping of Low-Statured Shrubland
Cover Types in Post-Agricultural Landscapes of the US Northeast.''}
\emph{International Journal of Remote Sensing} 43 (19-24): 7117--38.
\url{https://doi.org/10.1080/01431161.2022.2155086}.

\leavevmode\vadjust pre{\hypertarget{ref-Malmsheimer2008}{}}%
Malmsheimer, Robert W., Patrick Heffernan, Steve Brink, Douglas
Crandall, Fred Deneke, Christopher Galik, Edmund Gee, et al. 2008.
{``{Forest Management Solutions for Mitigating Climate Change in the
United States}.''} \emph{Journal of Forestry} 106 (3): 115--17.
\url{https://doi.org/10.1093/jof/106.3.115}.

\leavevmode\vadjust pre{\hypertarget{ref-Matasci2018}{}}%
Matasci, Giona, Txomin Hermosilla, Michael A. Wulder, Joanne C. White,
Nicholas C. Coops, Geordie W. Hobart, Douglas K. Bolton, Piotr
Tompalski, and Christopher W. Bater. 2018. {``Three Decades of Forest
Structural Dynamics over Canada{\textquotesingle}s Forested Ecosystems
Using Landsat Time-Series and Lidar Plots.''} \emph{Remote Sensing of
Environment} 216 (October): 697--714.
\url{https://doi.org/10.1016/j.rse.2018.07.024}.

\leavevmode\vadjust pre{\hypertarget{ref-McRoberts2011}{}}%
McRoberts, Ronald E. 2011. {``Satellite Image-Based Maps: Scientific
Inference or Pretty Pictures?''} \emph{Remote Sensing of Environment}
115 (2): 715--24. \url{https://doi.org/10.1016/j.rse.2010.10.013}.

\leavevmode\vadjust pre{\hypertarget{ref-Menlove2020}{}}%
Menlove, James, and Sean P. Healey. 2020. {``A Comprehensive Forest
Biomass Dataset for the USA Allows Customized Validation of Remotely
Sensed Biomass Estimates.''} \emph{Remote Sensing} 12 (24).
\url{https://doi.org/10.3390/rs12244141}.

\leavevmode\vadjust pre{\hypertarget{ref-MERENLENDER2004}{}}%
Merenlender, A. M., L. Huntsinger, G. Guthey, and S. K. Fairfax. 2004.
{``Land Trusts and Conservation Easements: Who Is Conserving What for
Whom?''} \emph{Conservation Biology} 18 (1): 65--76.
\url{https://doi.org/10.1111/j.1523-1739.2004.00401.x}.

\leavevmode\vadjust pre{\hypertarget{ref-Meyer2021}{}}%
Meyer, Hanna, and Edzer Pebesma. 2021. {``{Predicting into unknown
space? Estimating the area of applicability of spatial prediction
models}.''} \emph{Methods in Ecology and Evolution} 12 (9): 1620--33.
\url{https://doi.org/10.1111/2041-210x.13650}.

\leavevmode\vadjust pre{\hypertarget{ref-Montero2022}{}}%
Montero, David, Cesar Aybar, Miguel D. Mahecha, and Sebastian Wieneke.
2022. {``Spectral: Awesome Spectral Indices Deployed via the Google
Earth Engine JavaScript API.''} \emph{The International Archives of the
Photogrammetry, Remote Sensing and Spatial Information Sciences}
XLVIII-4/W1-2022: 301--6.
\url{https://doi.org/10.5194/isprs-archives-XLVIII-4-W1-2022-301-2022}.

\leavevmode\vadjust pre{\hypertarget{ref-Omernik2014}{}}%
Omernik, James M., and Glenn E. Griffith. 2014. {``Ecoregions of the
Conterminous United States: Evolution of a Hierarchical Spatial
Framework.''} \emph{Environmental Management} 54 (6): 1249--66.
\url{https://doi.org/10.1007/s00267-014-0364-1}.

\leavevmode\vadjust pre{\hypertarget{ref-highpeaks}{}}%
Pataki, George E, and John P Cahill. 1999. {``High Peaks Wilderness
Complex Unit Management Plan.''}

\leavevmode\vadjust pre{\hypertarget{ref-Patton2022}{}}%
Patton, Ry M, Diane H Kiernan, Julia I Burton, and John E Drake. 2022.
{``Management Trade-Offs Between Forest Carbon Stocks, Sequestration
Rates and Structural Complexity in the Central Adirondacks.''}
\emph{Forest Ecology and Management} 525: 120539.
\url{https://doi.org/10.1016/j.foreco.2022.120539}.

\leavevmode\vadjust pre{\hypertarget{ref-sf}{}}%
Pebesma, Edzer. 2018. {``{Simple Features for R: Standardized Support
for Spatial Vector Data}.''} \emph{{The R Journal}} 10 (1): 439--46.
\url{https://doi.org/10.32614/RJ-2018-009}.

\leavevmode\vadjust pre{\hypertarget{ref-Pengra2020}{}}%
Pengra, Bruce, Steve Stehman, Josephine A Horton, Daryn J Dockter, Todd
A Schroeder, Zhiqiang Yang, Alex J Hernandez, et al. 2020. {``LCMAP
Reference Data Product 1984-2018 Land Cover, Land Use and Change Process
Attributes (Ver. 1.2, November 2021).''}
\url{https://doi.org/10.5066/P9ZWOXJ7}.

\leavevmode\vadjust pre{\hypertarget{ref-PRISM}{}}%
PRISM Climate Group. 2022. {``PRISM Climate Data.''}
\url{https://prism.oregonstate.edu}.

\leavevmode\vadjust pre{\hypertarget{ref-BIOMASS_SAT}{}}%
Quegan, Shaun, Thuy Le Toan, Jerome Chave, Jorgen Dall, Jean-Francois
Exbrayat, Dinh Ho Tong Minh, Mark Lomas, et al. 2019. {``The European
Space Agency BIOMASS Mission: Measuring Forest Above-Ground Biomass from
Space.''} \emph{Remote Sensing of Environment} 227: 44--60.
\url{https://doi.org/10.1016/j.rse.2019.03.032}.

\leavevmode\vadjust pre{\hypertarget{ref-R}{}}%
R Core Team. 2021. \emph{R: A Language and Environment for Statistical
Computing}. Vienna, Austria: R Foundation for Statistical Computing.
\url{https://www.R-project.org/}.

\leavevmode\vadjust pre{\hypertarget{ref-Riemann2010}{}}%
Riemann, Rachel, Barry Tyler Wilson, Andrew Lister, and Sarah Parks.
2010. {``{An effective assessment protocol for continuous geospatial
datasets of forest characteristics using {USFS} Forest Inventory and
Analysis ({FIA}) data}.''} \emph{Remote Sensing of Environment} 114
(10): 2337--52. \url{https://doi.org/10.1016/j.rse.2010.05.010}.

\leavevmode\vadjust pre{\hypertarget{ref-ALOS}{}}%
Rosenqvist, A., M. Shimada, N. Ito, and M. Watanabe. 2007. {``{ALOS}
{PALSAR}: A Pathfinder Mission for Global-Scale Monitoring of the
Environment.''} \emph{{IEEE} Transactions on Geoscience and Remote
Sensing} 45 (11): 3307--16.
\url{https://doi.org/10.1109/tgrs.2007.901027}.

\leavevmode\vadjust pre{\hypertarget{ref-Roy2016}{}}%
Roy, D. P., V. Kovalskyy, H. K. Zhang, E. F. Vermote, L. Yan, S. S.
Kumar, and A. Egorov. 2016. {``Characterization of Landsat-7 to
Landsat-8 Reflective Wavelength and Normalized Difference Vegetation
Index Continuity.''} \emph{Remote Sensing of Environment} 185
(November): 57--70. \url{https://doi.org/10.1016/j.rse.2015.12.024}.

\leavevmode\vadjust pre{\hypertarget{ref-Saarela2016}{}}%
Saarela, Svetlana, Sören Holm, Anton Grafström, Sebastian Schnell, Erik
Næsset, Timothy G. Gregoire, Ross F. Nelson, and Göran Ståhl. 2016.
{``Hierarchical Model-Based Inference for Forest Inventory Utilizing
Three Sources of Information.''} \emph{Annals of Forest Science} 73 (4):
895--910. \url{https://doi.org/10.1007/s13595-016-0590-1}.

\leavevmode\vadjust pre{\hypertarget{ref-lightgbm}{}}%
Shi, Yu, Guolin Ke, Damien Soukhavong, James Lamb, Qi Meng, Thomas
Finley, Taifeng Wang, et al. 2022. \emph{Lightgbm: Light Gradient
Boosting Machine}. \url{https://github.com/Microsoft/LightGBM}.

\leavevmode\vadjust pre{\hypertarget{ref-Simard2011}{}}%
Simard, Marc, Naiara Pinto, Joshua B Fisher, and Alessandro Baccini.
2011. {``Mapping Forest Canopy Height Globally with Spaceborne Lidar.''}
\emph{Journal of Geophysical Research: Biogeosciences} 116 (G4).
\url{https://doi.org/10.1029/2011JG001708}.

\leavevmode\vadjust pre{\hypertarget{ref-Skowronksi2012}{}}%
Skowronski, Nicholas S, and Andrew J Lister. 2012. {``{Utility of LiDAR
for large area forest inventory applications}.''} In \emph{In: Morin,
Randall s.; Liknes, Greg c., Comps. Moving from Status to Trends: Forest
Inventory and Analysis (FIA) Symposium 2012; 2012 December 4-6;
Baltimore, MD. Gen. Tech. Rep. NRS-p-105. Newtown Square, PA: US
Department of Agriculture, Forest Service, Northern Research
Station.{[}CD-ROM{]}: 410-413.}, 410--13.
\url{https://www.fs.usda.gov/treesearch/pubs/42792}.

\leavevmode\vadjust pre{\hypertarget{ref-Stehman2019}{}}%
Stehman, Stephen V., and Giles M. Foody. 2019. {``Key Issues in Rigorous
Accuracy Assessment of Land Cover Products.''} \emph{Remote Sensing of
Environment} 231 (September): 111199.
\url{https://doi.org/10.1016/j.rse.2019.05.018}.

\leavevmode\vadjust pre{\hypertarget{ref-Strunk2014}{}}%
Strunk, Jacob L., Hailemariam Temesgen, Hans-Erik Andersen, and Petteri
Packalen. 2014. {``Prediction of Forest Attributes with Field Plots,
Landsat, and a Sample of Lidar Strips.''} \emph{Photogrammetric
Engineering \& Remote Sensing} 80 (2): 143--50.
\url{https://doi.org/10.14358/pers.80.2.143-150}.

\leavevmode\vadjust pre{\hypertarget{ref-Sugarbaker2014}{}}%
Sugarbaker, Larry J., Eric W. Constance, Hans Karl Heidemann, Allyson L.
Jason, Vicki Lukas, David L. Saghy, and Jason M. Stoker. 2014. {``The 3D
Elevation Program Initiative: A Call for Action.''} {US} Geological
Survey. \url{https://doi.org/10.3133/cir1399}.

\leavevmode\vadjust pre{\hypertarget{ref-Sugarbaker2017}{}}%
Sugarbaker, Larry J., Diane F. Eldridge, Allyson L. Jason, Vicki Lukas,
David L. Saghy, Jason M. Stoker, and Diana R. Thunen. 2017. {``Status of
the 3D Elevation Program, 2015.''} {US} Geological Survey.
\url{https://doi.org/10.3133/ofr20161196}.

\leavevmode\vadjust pre{\hypertarget{ref-Sentinel-1}{}}%
Torres, Ramon, Paul Snoeij, Dirk Geudtner, David Bibby, Malcolm
Davidson, Evert Attema, Pierre Potin, et al. 2012. {``{GMES} Sentinel-1
Mission.''} \emph{Remote Sensing of Environment} 120 (May): 9--24.
\url{https://doi.org/10.1016/j.rse.2011.05.028}.

\leavevmode\vadjust pre{\hypertarget{ref-Urbazaev2018}{}}%
Urbazaev, Mikhail, Christian Thiel, Felix Cremer, Ralph Dubayah, Mirco
Migliavacca, Markus Reichstein, and Christiane Schmullius. 2018.
{``Estimation of Forest Aboveground Biomass and Uncertainties by
Integration of Field Measurements, Airborne {LiDAR}, and {SAR} and
Optical Satellite Data in Mexico.''} \emph{Carbon Balance and
Management} 13 (1). \url{https://doi.org/10.1186/s13021-018-0093-5}.

\leavevmode\vadjust pre{\hypertarget{ref-USGS_dem}{}}%
U.S. Geological Survey. 2019. {``{3D} Elevation Program 1-Meter
Resolution Digital Elevation Model.''}
\url{https://www.usgs.gov/the-national-map-data-delivery}.

\leavevmode\vadjust pre{\hypertarget{ref-Census}{}}%
US Census Bureau. 2013. {``TIGER/Line Shapefiles.''}
\url{https://www.census.gov/geographies/mapping-files/time-series/geo/tiger-line-file.html}.

\leavevmode\vadjust pre{\hypertarget{ref-insects2006}{}}%
USFS. 2006. \emph{Forest Insect and Disease Conditions in the United
States 2006}. USDA Forest Service.
\url{https://www.fs.usda.gov/foresthealth/publications/ConditionsReport_2006.pdf}.

\leavevmode\vadjust pre{\hypertarget{ref-NYForests2019}{}}%
---------. 2020. {``Forests of New York, 2019.''} United States
Department of Agriculture, Forest Service; U.S. Department of
Agriculture, Forest Service, Northern Research Station.
\url{https://doi.org/10.2737/fs-ru-250}.

\leavevmode\vadjust pre{\hypertarget{ref-LandsatC1}{}}%
USGS. 2018. {``Landsat Collections.''} {US} Geological Survey.
\url{https://doi.org/10.3133/fs20183049}.

\leavevmode\vadjust pre{\hypertarget{ref-Tigris}{}}%
Walker, Kyle. 2022. \emph{Tigris: Load Census TIGER/Line Shapefiles}.
\url{https://CRAN.R-project.org/package=tigris}.

\leavevmode\vadjust pre{\hypertarget{ref-White2017}{}}%
White, Joanne C., Michael A. Wulder, Txomin Hermosilla, Nicholas C.
Coops, and Geordie W. Hobart. 2017. {``A Nationwide Annual
Characterization of 25 Years of Forest Disturbance and Recovery for
Canada Using Landsat Time Series.''} \emph{Remote Sensing of
Environment} 194 (June): 303--21.
\url{https://doi.org/10.1016/j.rse.2017.03.035}.

\leavevmode\vadjust pre{\hypertarget{ref-Whitney1994}{}}%
Whitney, Gordon G. 1994. \emph{From Coastal Wilderness to Fruited Plain:
A History of Environmental Change in Temperate North America from 1500
to the Present}. Cambridge, United Kingdom: Cambridge University Press.

\leavevmode\vadjust pre{\hypertarget{ref-NYForests2015}{}}%
Widmann, Richard H. 2016. {``Forests of New York, 2015.''} United States
Department of Agriculture, Forest Service; U.S. Department of
Agriculture, Forest Service, Northern Research Station.
\url{https://doi.org/10.2737/fs-ru-96}.

\leavevmode\vadjust pre{\hypertarget{ref-NYForests2007}{}}%
Widmann, Richard H., Sloane Crawford, Charles Barnett, Brett J. Butler,
Grant M. Domke, Douglas M. Griffith, Mark A. Hatfield, et al. 2012.
{``New York{\textquotesingle}s Forests 2007.''} United States Department
of Agriculture, Forest Service; U.S. Department of Agriculture, Forest
Service, Northern Research Station.
\url{https://doi.org/10.2737/nrs-rb-65}.

\leavevmode\vadjust pre{\hypertarget{ref-Wilen1995}{}}%
Wilen, B. O., and M. K. Bates. 1995. {``The {US} Fish and Wildlife
Service's National Wetlands Inventory Project.''} In
\emph{Classification and Inventory of the World's Wetlands}, 153--69.
Springer Netherlands.
\url{https://doi.org/10.1007/978-94-011-0427-2_13}.

\leavevmode\vadjust pre{\hypertarget{ref-Wintle2003}{}}%
Wintle, B. A., M. A. McCarthy, C. T. Volinksy, and R. P. Kavanagh. 2003.
{``The Use of Bayesian Model Averaging to Better Represent Uncertainty
in Ecological Models.''} \emph{Conservation Biology} 17 (6): 1579--90.
\url{https://doi.org/10.1111/j.1523-1739.2003.00614.x}.

\leavevmode\vadjust pre{\hypertarget{ref-Wolpert1992}{}}%
Wolpert, David H. 1992. {``Stacked Generalization.''} \emph{Neural
Networks} 5 (2): 241--59.
\url{https://doi.org/10.1016/S0893-6080(05)80023-1}.

\leavevmode\vadjust pre{\hypertarget{ref-Woodall2015}{}}%
Woodall, Christopher W., John W. Coulston, Grant M. Domke, Brian F.
Walters, David N. Wear, James E. Smith, Hans-Erik Andersen, et al. 2015.
{``The u.s. Forest Carbon Accounting Framework: Stocks and Stock Change,
1990-2016.''} U.S. Department of Agriculture, Forest Service, Northern
Research Station. \url{https://doi.org/10.2737/nrs-gtr-154}.

\leavevmode\vadjust pre{\hypertarget{ref-Woodall2011}{}}%
Woodall, Christopher W., Linda S. Heath, Grant M. Domke, and Michael C.
Nichols. 2011. {``Methods and Equations for Estimating Aboveground
Volume, Biomass, and Carbon for Trees in the u.s. Forest Inventory,
2010.''} U.S. Department of Agriculture, Forest Service, Northern
Research Station. \url{https://doi.org/10.2737/nrs-gtr-88}.

\leavevmode\vadjust pre{\hypertarget{ref-ranger}{}}%
Wright, Marvin N., and Andreas Ziegler. 2017. {``{ranger}: A Fast
Implementation of Random Forests for High Dimensional Data in {C++} and
{R}.''} \emph{Journal of Statistical Software} 77 (1): 1--17.
\url{https://doi.org/10.18637/jss.v077.i01}.

\leavevmode\vadjust pre{\hypertarget{ref-Landsat50}{}}%
Wulder, Michael A., David P. Roy, Volker C. Radeloff, Thomas R.
Loveland, Martha C. Anderson, David M. Johnson, Sean Healey, et al.
2022. {``Fifty Years of Landsat Science and Impacts.''} \emph{Remote
Sensing of Environment} 280: 113195.
\url{https://doi.org/10.1016/j.rse.2022.113195}.

\leavevmode\vadjust pre{\hypertarget{ref-Wulder2012}{}}%
Wulder, Michael A., Joanne C. White, Ross F. Nelson, Erik Næsset, Hans
Ole Ørka, Nicholas C. Coops, Thomas Hilker, Christopher W. Bater, and
Terje Gobakken. 2012. {``Lidar Sampling for Large-Area Forest
Characterization: A Review.''} \emph{Remote Sensing of Environment} 121
(June): 196--209. \url{https://doi.org/10.1016/j.rse.2012.02.001}.

\leavevmode\vadjust pre{\hypertarget{ref-Zhu2014}{}}%
Zhu, Zhe, and Curtis E. Woodcock. 2014. {``{Continuous change detection
and classification of land cover using all available Landsat data}.''}
\emph{Remote Sensing of Environment} 144: 152--71.
\url{https://doi.org/10.1016/j.rse.2014.01.011}.

\end{CSLReferences}

\end{document}


\maketitle
\ifdefined\Shaded\renewenvironment{Shaded}{\begin{tcolorbox}[boxrule=0pt, breakable, enhanced, interior hidden, borderline west={3pt}{0pt}{shadecolor}, frame hidden, sharp corners]}{\end{tcolorbox}}\fi

\renewcommand*\contentsname{Table of contents}
{
\hypersetup{linkcolor=}
\setcounter{tocdepth}{3}
\tableofcontents
}
\newpage{}

\hypertarget{supplementary-materials-1-landtrendr-parameters}{%
\subsection{Supplementary Materials 1: Landtrendr
parameters}\label{supplementary-materials-1-landtrendr-parameters}}

\FloatBarrier{}

\begin{longtable}[]{@{}
  >{\raggedright\arraybackslash}p{(\columnwidth - 4\tabcolsep) * \real{0.2527}}
  >{\raggedleft\arraybackslash}p{(\columnwidth - 4\tabcolsep) * \real{0.4945}}
  >{\raggedleft\arraybackslash}p{(\columnwidth - 4\tabcolsep) * \real{0.2527}}@{}}
\caption{Landtrendr Google Earth Engine (LT-GEE) segmentation parameters
for 16 Landsat-derived predictors (Kennedy et al. 2018).
}\tabularnewline
\toprule()
\begin{minipage}[b]{\linewidth}\raggedright
Parameters
\end{minipage} & \begin{minipage}[b]{\linewidth}\raggedleft
Annual Reflectance (NBR, NDVI, SR, MSR, TC*)
\end{minipage} & \begin{minipage}[b]{\linewidth}\raggedleft
Disturbance (YOD, MAG)
\end{minipage} \\
\midrule()
\endfirsthead
\toprule()
\begin{minipage}[b]{\linewidth}\raggedright
Parameters
\end{minipage} & \begin{minipage}[b]{\linewidth}\raggedleft
Annual Reflectance (NBR, NDVI, SR, MSR, TC*)
\end{minipage} & \begin{minipage}[b]{\linewidth}\raggedleft
Disturbance (YOD, MAG)
\end{minipage} \\
\midrule()
\endhead
maxSegments & 5 & 10 \\
spikeThreshold & 0.5 & 0.9 \\
vertexCountOvershoot & 3 & 3 \\
preventOneYearRecovery & true & true \\
recoveryThreshold & 0.25 & 0.75 \\
pvalThreshold & 0.05 & 0.05 \\
bestModelProportion & 0.75 & 0.75 \\
minObservationsNeeded & 6 & 6 \\
\bottomrule()
\end{longtable}

\begin{table}[H]
\caption{Disturbance predictor (YOD, MAG) parameters in LT-GEE.}\tabularnewline

\centering
\begin{tabular}[t]{ccc}
\toprule
\multicolumn{1}{c}{Parameter} & \multicolumn{1}{c}{Value} & \multicolumn{1}{c}{Operator}\\
\midrule
Delta & Loss & \\
\addlinespace
Sort & Most recent & \\
\addlinespace
Year & 1985-Target year & \\
\addlinespace
Magnitude & 50 & Greater than\\
\addlinespace
Duration & 4 & Less than\\
\addlinespace
Pre-disturbance spectral value & 300 & Greater than\\
\addlinespace
Minimum mapping unit & 7 & \\
\bottomrule
\end{tabular}
\end{table}

\FloatBarrier{}

\newpage{}

\hypertarget{supplementary-materials-2-model-development}{%
\subsection{Supplementary Materials 2: Model
Development}\label{supplementary-materials-2-model-development}}

\FloatBarrier{}

\begin{figure}

{\centering \includegraphics{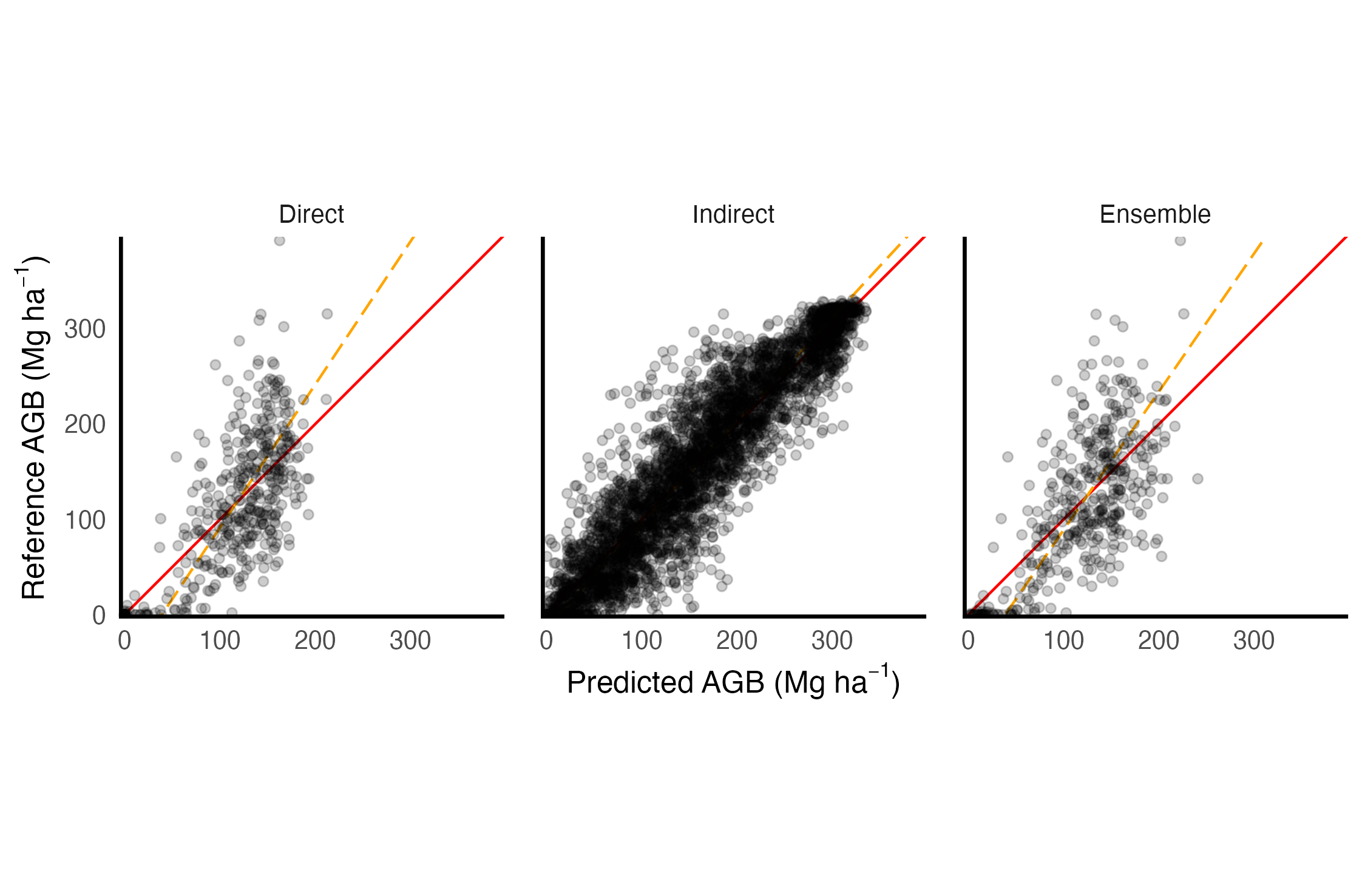}

}

\caption{\label{fig-testscatter}Reference vs predicted AGB scatter plot
for the 20\% testing portion of the model datasets (direct n = 410;
Indirect n = 4000; Ensemble n = 410). AGB values in Mg
ha\textsuperscript{-1} . Geometric mean functional relationship (GMFR)
trend line shown with dashed (orange) line, and 1:1 line shown with
solid (red) line.}

\end{figure}

\hypertarget{tbl-test}{}
\begin{table}
\caption{\label{tbl-test}Model performance metrics (as defined in Section 2.6) against 20\%
testing partition of the model datasets (Direct n = 631; Indirect n =
4000; Ensemble n = 631). }\tabularnewline

\centering
\begin{tabular}[t]{lrrr}
\toprule
\multicolumn{1}{c}{ } & \multicolumn{1}{c}{Direct} & \multicolumn{1}{c}{Indirect} & \multicolumn{1}{c}{Ensemble}\\
\midrule
RMSE & 52.50 & 36.91 & 53.37\\
\addlinespace
\% RMSE & 44.19 & 22.28 & 44.92\\
\addlinespace
MAE & 40.23 & 27.51 & 41.75\\
\addlinespace
\% MAE & 33.86 & 16.61 & 35.14\\
\addlinespace
ME & -0.54 & -0.46 & 1.70\\
\addlinespace
R² & 0.48 & 0.85 & 0.46\\
\bottomrule
\end{tabular}
\end{table}

\FloatBarrier{}

\begin{table}[H]
\caption{Selected hyperparemeters for final random forests (RF).}\tabularnewline

\centering\begingroup\fontsize{10}{12}\selectfont

\begin{tabular}[t]{lrrrrl}
\toprule
\multicolumn{1}{c}{Model} & \multicolumn{1}{c}{Num Trees} & \multicolumn{1}{c}{Mtry} & \multicolumn{1}{c}{Min Node Size} & \multicolumn{1}{c}{Sample Fraction} & \multicolumn{1}{c}{Replace}\\
\midrule
Direct & 1,500 & 31 & 2 & 0.85 & True\\
\addlinespace
Indirect & 1,000 & 13 & 3 & 1 & False\\
\bottomrule
\end{tabular}
\endgroup{}
\end{table}

\begin{table}[H]
\caption{Selected hyperparemeters for final stochastic gradient boosting machines
(LGB).}\tabularnewline

\centering\begingroup\fontsize{10}{12}\selectfont

\resizebox{\linewidth}{!}{
\begin{tabular}[t]{lrrrrlrrrrrrr}
\toprule
\multicolumn{1}{c}{Model} & \multicolumn{1}{c}{Learning Rate} & \multicolumn{1}{c}{Num Rounds} & \multicolumn{1}{c}{Num Leaves} & \multicolumn{1}{c}{Max Depth} & \multicolumn{1}{c}{Extra Trees} & \multicolumn{1}{c}{Min Data In Leaf} & \multicolumn{1}{c}{Bagging Frac} & \multicolumn{1}{c}{Bagging Freq} & \multicolumn{1}{c}{Feature Frac} & \multicolumn{1}{c}{Min Data In Bin} & \multicolumn{1}{c}{L1} & \multicolumn{1}{c}{L2}\\
\midrule
Direct & 0.05 & 100 & 16 & 24 & True & 16 & 0.8 & 6 & 0.8 & 14 & 0.1 & 0.1\\
\addlinespace
Indirect & 0.10 & 4,000 & 43 & 24 & True & 3 & 1.0 & 5 & 0.7 & 15 & 8.0 & 5.0\\
\bottomrule
\end{tabular}}
\endgroup{}
\end{table}

\begin{table}[H]
\caption{Selected hyperparemeters for final support vector machine (SVM).}\tabularnewline

\centering\begingroup\fontsize{10}{12}\selectfont

\begin{tabular}[t]{lllrrr}
\toprule
\multicolumn{1}{c}{Model} & \multicolumn{1}{c}{Kernel} & \multicolumn{1}{c}{Type} & \multicolumn{1}{c}{Sigma} & \multicolumn{1}{c}{C} & \multicolumn{1}{c}{Epsilon}\\
\midrule
Direct & Laplacian & Epsilon Support Vector Regression & 0.0019531 & 36 & 0.0441942\\
\bottomrule
\end{tabular}
\endgroup{}
\end{table}

\begin{longtable}[]{@{}lr@{}}
\caption{Direct linear model ensemble. Coefficients, rounded to the
nearest thousandth place, for each component model prediction.
}\tabularnewline
\toprule()
Term & Coefficient \\
\midrule()
\endfirsthead
\toprule()
Term & Coefficient \\
\midrule()
\endhead
Intercept & -12.223 \\
RF & 0.733 \\
LGB & 0.091 \\
SVM & 0.277 \\
\bottomrule()
\end{longtable}

\newpage{}

\begin{longtable}[]{@{}lr@{}}
\caption{Indirect linear model ensemble. Coefficients, rounded to the
nearest thousandth place, for each component model prediction.
}\tabularnewline
\toprule()
Term & Coefficient \\
\midrule()
\endfirsthead
\toprule()
Term & Coefficient \\
\midrule()
\endhead
Intercept & -4.067 \\
RF & 0.335 \\
LGB & 0.688 \\
\bottomrule()
\end{longtable}

\newpage{}

\FloatBarrier{}

\hypertarget{supplementary-materials-3-accuracyagreement-metrics}{%
\subsection{Supplementary Materials 3: Accuracy/Agreement
Metrics}\label{supplementary-materials-3-accuracyagreement-metrics}}

\begin{equation}\protect\hypertarget{eq-rmse}{}{
\operatorname{RMSE} = \sqrt{(\frac{1}{n})\sum_{i=1}^{n}(y_{i} - \hat{y_{i}})^{2}} 
}\label{eq-rmse}\end{equation}

\begin{equation}\protect\hypertarget{eq-prmse}{}{
\operatorname{\%\ RMSE} = 100 \cdot \frac{\operatorname{RMSE}}{\bar{y}}
}\label{eq-prmse}\end{equation}

\begin{equation}\protect\hypertarget{eq-mae}{}{
\operatorname{MAE} =  (\frac{1}{n})\sum_{i=1}^{n}(\lvert y_{i} - \hat{y_{i}} \rvert)
}\label{eq-mae}\end{equation}

\begin{equation}\protect\hypertarget{eq-pmae}{}{
\operatorname{\%\ MAE} = 100 \cdot \frac{\operatorname{MAE}}{\bar{y}}
}\label{eq-pmae}\end{equation}

\begin{equation}\protect\hypertarget{eq-me}{}{
\operatorname{ME} = (\frac{1}{n})\sum_{i=1}^{n}(y_{i} - \hat{y_{i}})
}\label{eq-me}\end{equation}

\begin{equation}\protect\hypertarget{eq-r2}{}{
\operatorname{R^2} = 1 - \frac{\sum_{i=1}^{n}\left(y_{i}-\hat{y}_{i}\right)^2}{\sum_{i=1}^{n}\left(y_i - \bar{y}\right)^2}
}\label{eq-r2}\end{equation}

\noindent where \(n\) is the number of reference obsevations included in
the data set, \(\hat{y_i}\) is the predicted value of AGB, \(y_{i}\) the
the corresponding reference AGB value, and \(\bar{y}\) the mean
reference AGB value.

Standard errors for R\textsuperscript{2} and RMSE were computed as
follows: \begin{equation}\protect\hypertarget{eq-bootse}{}{
\operatorname{SE_{boot}} = \sqrt{\frac{Var_{boot}}{n}} 
}\label{eq-bootse}\end{equation}

\noindent where \(n\) is the number of reference observations included
in the dataset, and \(Var_{boot}\) is computed as the variance of
R\textsuperscript{2} and RMSE estimates for 1000 iterations of bootstrap
resampling. Standard errors for MAE and ME were computed as follows:

\begin{equation}\protect\hypertarget{eq-se}{}{
\operatorname{SE} = \sqrt{\frac{\sum_{i=1}^{n}\left(e_i - \bar{e}\right)^2}{n - 1}}
}\label{eq-se}\end{equation}

\noindent where \(e_{i}\) is the error for a given reference
observation, and \(\bar{e}\) is the mean error from all reference
observations included. In the case of MAE,
\(e_{i} = |y_{i} - \hat{y_{i}}|\), the absolute value of the error for a
given reference observation.

\newpage{}

\FloatBarrier{}

\hypertarget{supplementary-materials-4-map-agreement}{%
\subsection{Supplementary Materials 4: Map
Agreement}\label{supplementary-materials-4-map-agreement}}

\begin{figure}

{\centering \includegraphics{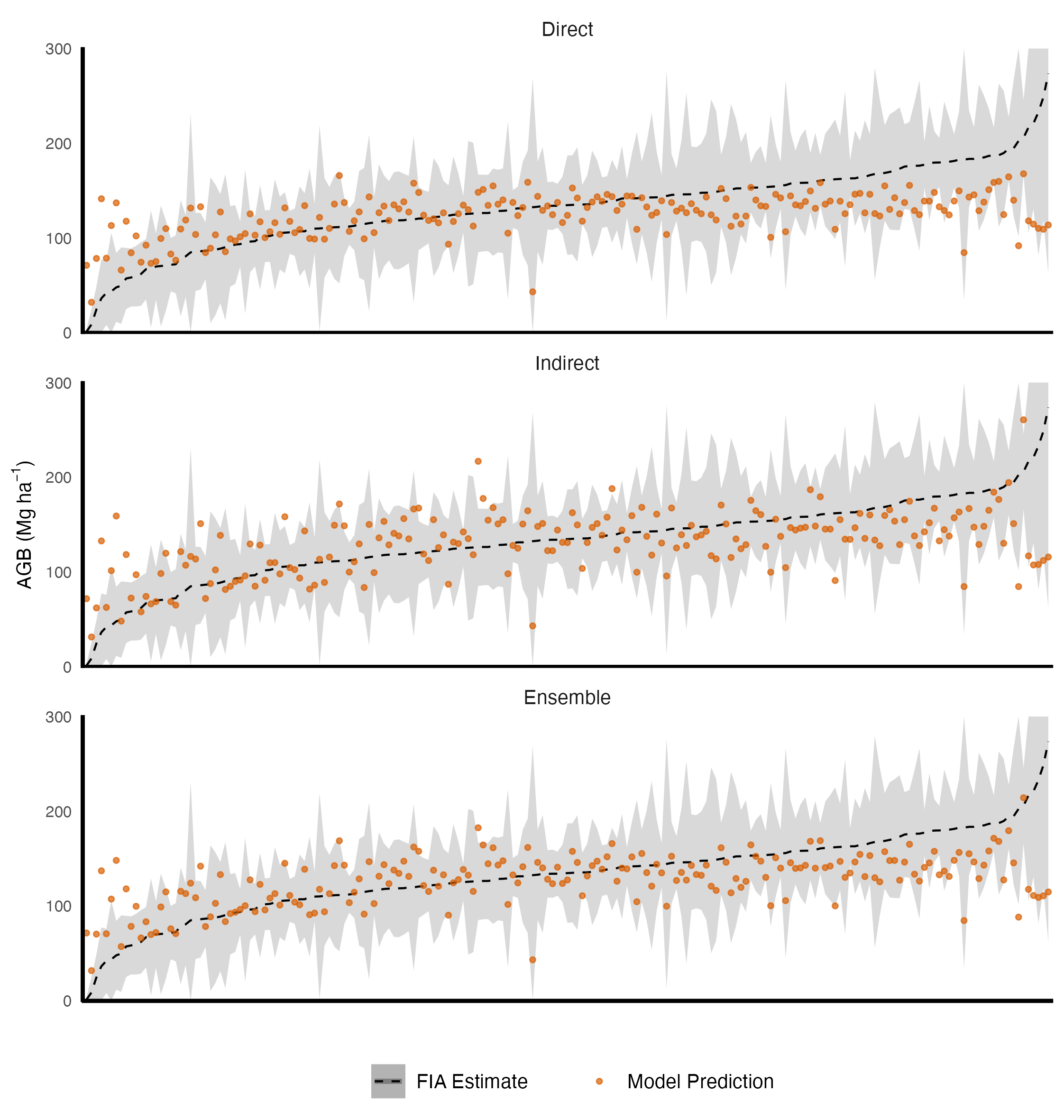}

}

\caption{\label{fig-menlove-healey}Comparison of 2016 AGB predictions to
(Menlove and Healey 2020) estimates (dashed line) and associated 95\%
confidence interval (gray shaded region) within 64,000-ha aggregation
hexagons. FIA estimates are scaled by the proportion of forest cover
indicated by LCMAP 2016 Tree cover, Wetland, and Grass/Shrub classified
pixels. Hexagons with \(<\) 50\% of their total area outside of New York
State's borders were excluded from this analysis. One~large outlier
(466.51 Mg ha\textsuperscript{-1}) was excluded, and upper CI boundaries
were capped at 300 Mg ha\textsuperscript{-1} for display purposes.
Observations are sorted by increasing FIA estimates along the x-axis.}

\end{figure}

\newpage{}

\begin{figure}

{\centering \includegraphics{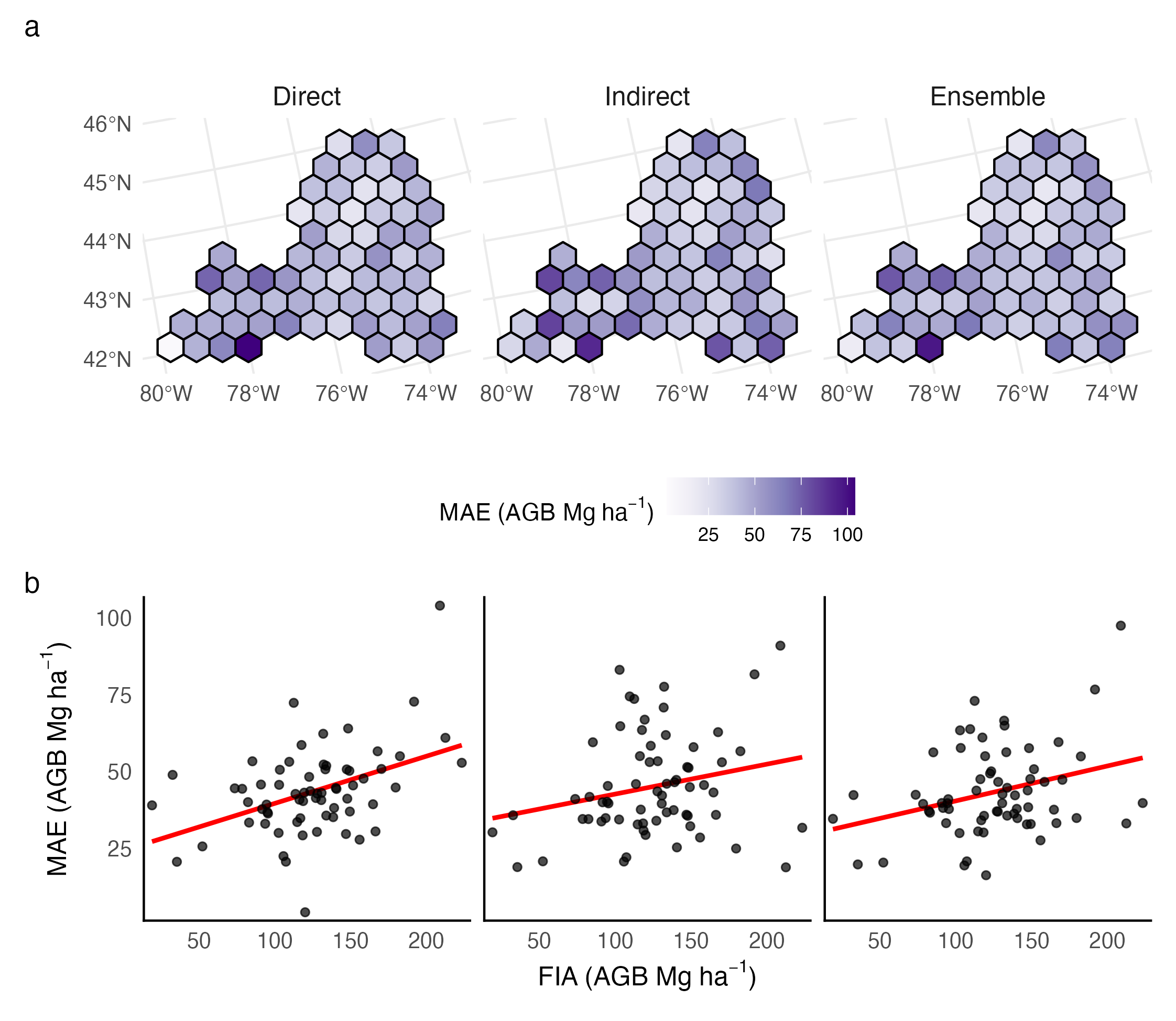}

}

\caption{\label{fig-mae}Spatial mean absolute error (MAE). a) MAE (Mg
ha\(^{-1}\)) computed from plot-to-pixel residuals summarized at units
spaced 50 km apart. Hexagons with only one reference plot were removed.
b) Hex-level MAE as a function of mean FIA AGB. Trend lines produced
using ordinary least squares regression.}

\end{figure}

\newpage{}

\begin{figure}

{\centering \includegraphics{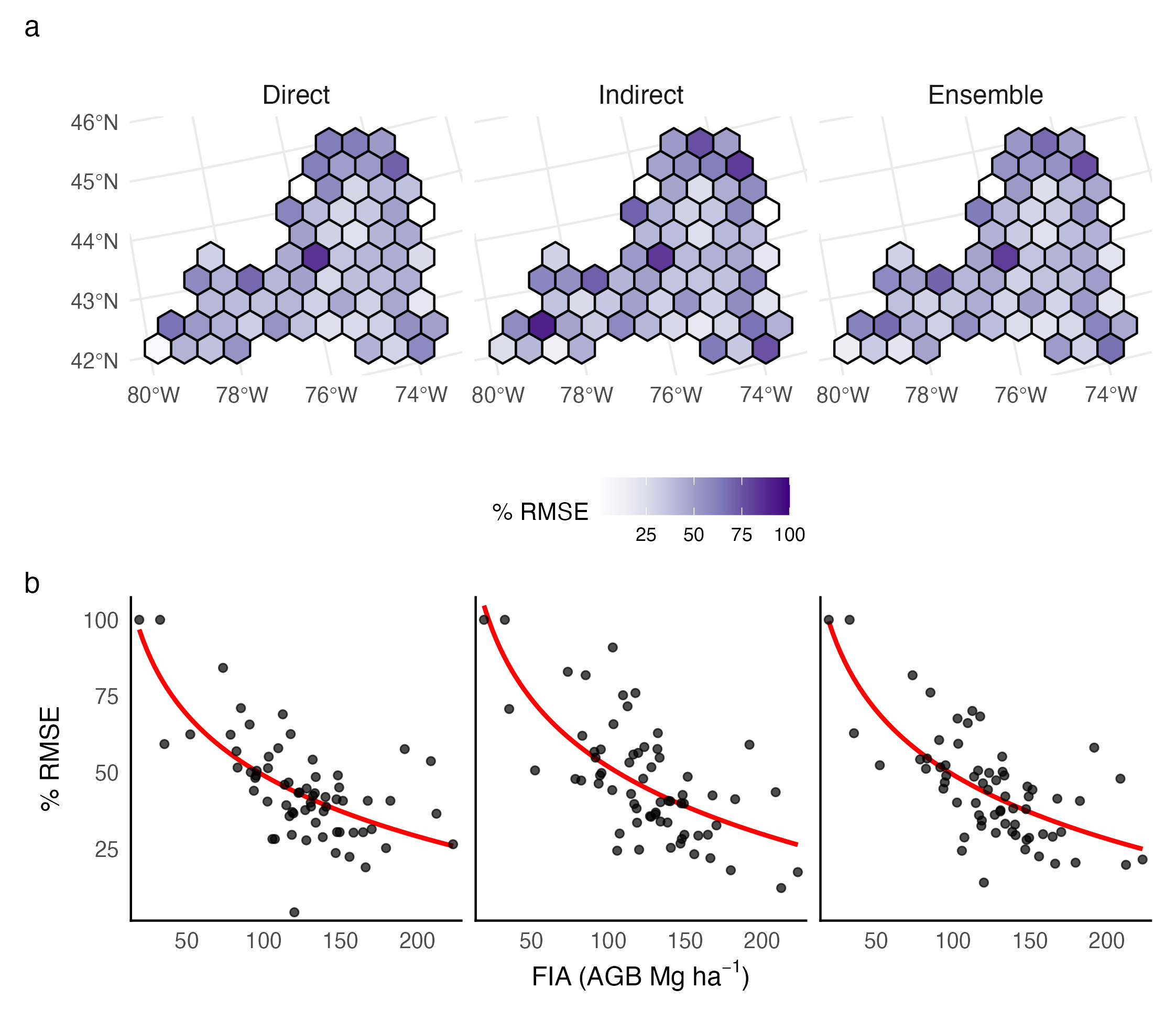}

}

\caption{\label{fig-rmse}Spatial \% root mean squared error (RMSE). a)
\% RMSE computed from plot-to-pixel residuals summarized at units spaced
50 km apart. Hexagons with only one reference plot were removed. b)
Hex-level \% RMSE as a function of mean FIA AGB. Trend lines produced
using logarithmic regression. \% RMSE values in a) and b) capped at 100
for display purposes.}

\end{figure}

\newpage{}

\begin{figure}

{\centering \includegraphics{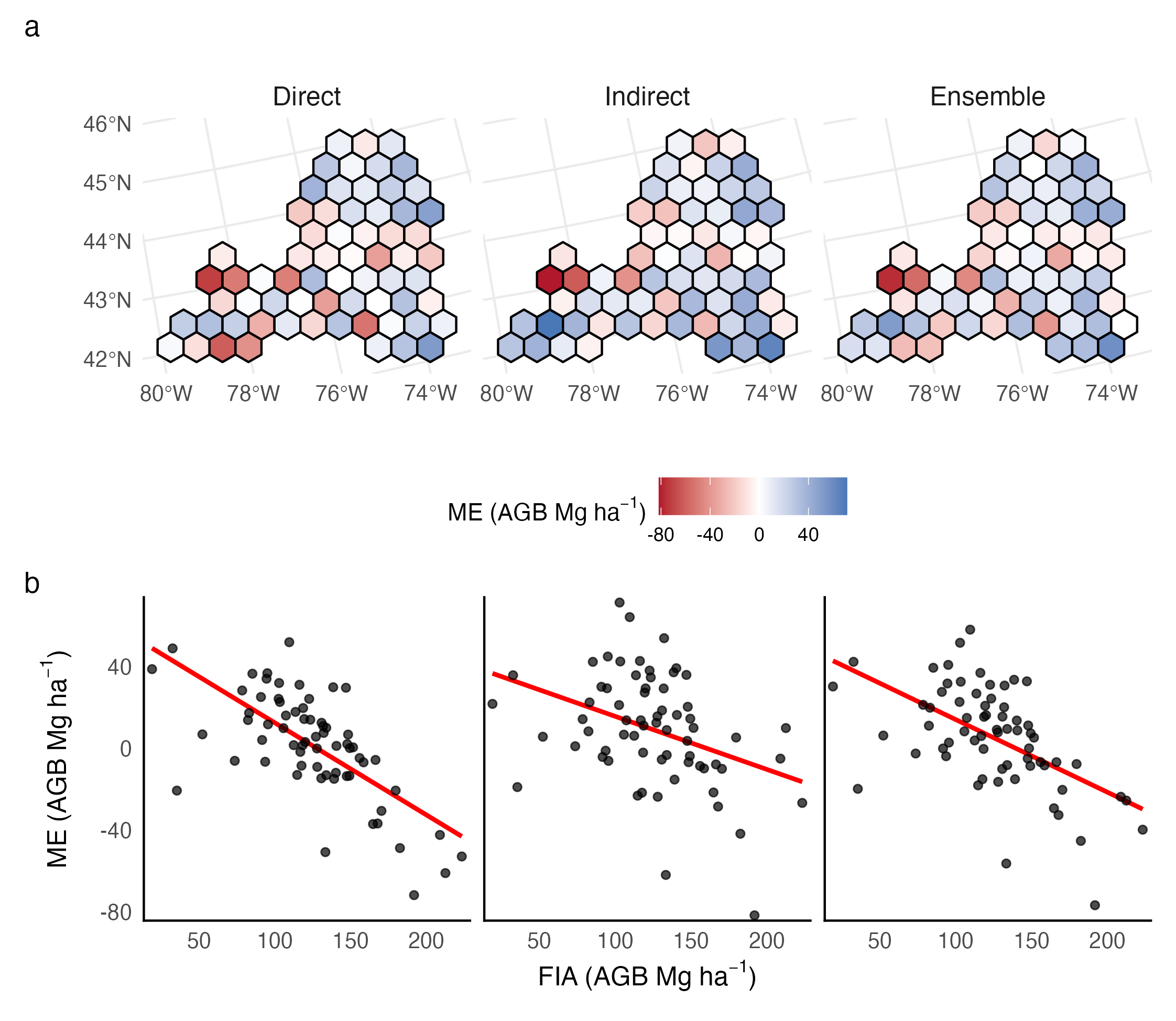}

}

\caption{\label{fig-me}Spatial mean error (ME). a) ME (Mg ha\(^{-1}\))
computed from plot-to-pixel residuals summarized at units spaced 50 km
apart. Hexagons with only one reference plot were removed. b) Hex-level
ME as a function of mean FIA AGB. Trend lines produced using ordinary
least squares regression.}

\end{figure}

\newpage{}

\FloatBarrier{}

\hypertarget{references}{%
\subsection*{References}\label{references}}
\addcontentsline{toc}{subsection}{References}

\hypertarget{refs}{}
\begin{CSLReferences}{1}{0}
\leavevmode\vadjust pre{\hypertarget{ref-Menlove2020}{}}%
Menlove, James, and Sean P. Healey. 2020. {``A Comprehensive Forest
Biomass Dataset for the USA Allows Customized Validation of Remotely
Sensed Biomass Estimates.''} \emph{Remote Sensing} 12 (24).
\url{https://doi.org/10.3390/rs12244141}.

\end{CSLReferences}